\documentclass[journal,twoside]{IEEEtran}

\usepackage{cite}
\usepackage{amsmath,amssymb,amsfonts}
\usepackage{algorithmic}
\usepackage{graphicx,color}
\usepackage{textcomp}
\usepackage{xcolor}
\usepackage{hyperref}
\hypersetup{hidelinks=true}
\usepackage{algorithm,algorithmic}
\def\BibTeX{{\rm B\kern-.05em{\sc i\kern-.025em b}\kern-.08em 
    T\kern-.1667em\lower.7ex\hbox{E}\kern-.125emX}}
\AtBeginDocument{\definecolor{tmlcncolor}{cmyk}{0.93,0.59,0.15,0.02}\definecolor{NavyBlue}{RGB}{0,86,125}}

\newcommand{\norm}[1]{\left\lVert#1\right\rVert}

\def\x{\mathbf{x}(t,k)}
\def\sq{s_q(t,k)}

\def\hq{\mathbf{h}_q(t,k)}

\def\hn{\mathbf{h}_n(t,k)}

\definecolor{revgreen}{RGB}{0,0,0}
\newcommand{\rev}[1]{\textcolor{revgreen}{#1}}

\usepackage{subcaption}
\usepackage{cite}
\usepackage{graphicx}
\usepackage{comment}
\usepackage{balance}
\usepackage{acronym}
\usepackage{amsmath,amssymb}
\usepackage{tabularx}
\usepackage{multirow}

\usepackage{booktabs}
\usepackage{tikz}
\usetikzlibrary{positioning}
\usepackage{caption}
\usetikzlibrary{shapes.multipart}
\usepackage{amsfonts}
\usepackage{bbm}
\usepackage{wrapfig}
\usepackage[normalem]{ulem}
\usepackage[utf8]{inputenc}
\acrodef{STFT}{short-time Fourier transform}
\acrodef{ISTFT}{inverse short-time Fourier transform}
\acrodef{BSS}{blind source separation}
\acrodef{DOA}{direction of arrival}
\acrodef{DC}{deep clustering}
\acrodef{DPRNN}{dual-path recurrent neural network}
\acrodef{TF}{time-frequency}
\acrodef{TCN}{temporal convolutional network}
\acrodef{ATF}{acoustic transfer function}
\acrodef{SI-SDR}{scale-invariant signal-to-distortion ratio}
\acrodef{MSE}{mean square error}
\acrodef{DFT}{discrete Fourier transform }
\acrodef{SIR}{signal-to-interference ratio}
\acrodef{OVA}{overlap-and-add}
\acrodef{SDR}{signal-to-distortion ratio}
\acrodef{BLSTM}{Bidirectional Long Short-Term Memory}
\acrodef{SOTA}{state-of-the-art}
\acrodef{RI}{Real-Imaginary}
\acrodef{RIR}{room impulse response}
\acrodef{SNR}{signal-to-noise ratio}
\acrodef{RNN}{Recurrent Neural Networks}
\acrodef{FLOP}{floating point operation}
\acrodef{WPE}{weighted prediction error}
\acrodef{STOA}{state of the art}
\acrodef{FC}{fully connected}
\acrodef{E2E}{end-to-end}
\acrodef{SDR}{signal to distortion ration}
\acrodef{SIR}{signal-to-interference ratio}
\acrodef{STOI}{short-time objective intelligibility}
\acrodef{PESQ}{perceptual evaluation of speech quality}
\acrodef{CNN}{convolutional neural network}
\acrodef{MVDR}{minimum variance distortion beamformer}
\acrodef{RTF}{relative transfer function}
\acrodef{IPD}{inter-channel phase difference}
\acrodef{LPS}{logarithm power spectra}
\acrodef{AF}{angle features}
\acrodef{DSB}{delay-and-sum beamformer}
\acrodef{SI-SDR}{scale-invariant signal-to-distortion ratio}
\acrodef{FiLM}{feature-wise linear modulation}
\acrodef{ULA}{uniform linear array}
\acrodef{TPD}{target phase differenc}
\acrodef{SISDR}{scale-invariant signal-to-distortion ratio}
\acrodef{TSE}{target speaker extraction}
\acrodef{PReLU}{parametric rectified linear unit}
\acrodef{RPC}{Relative Position Cues}
\acrodef{FiLM}{feature-wise linear modulation}
\acrodef{PIT}{permutation invariant training}
\acrodef{SSF}{spatially selective filter}

\tikzset{
  block/.style={
    rectangle,
    rounded corners,
    draw=black,
    very thick,
    minimum height=3em,
    minimum width=3em
  }
}
\tikzset{
  input/.style={
    circle,
    draw=black,
    very thick,
    minimum size=2em
  }
}
\tikzset{
  output/.style={
    circle,
    draw=black,
    very thick,
    minimum size=2em
  }
}

\begin{document}

\markboth{IEEE Transactions on Audio, Speech and Language Processing, Nov. 2026}%
{Eisenberg \MakeLowercase{\textit{et al.}}: Spatially-Guided Speaker Extraction Based on Relative Position Cues}

\title{Spatially-Guided Speaker Extraction Based on Relative Position Cues\thanks{This work was partially supported by the Israel Science Foundation (ISF) and the German Research Foundation (DFG) through the ISF–DFG Joint Research Program, Grant No.~1280/25.}
}

\author{Aviad Eisenberg, 
Shlomo E. Chazan~\IEEEmembership{Member,~IEEE,} and Sharon Gannot~\IEEEmembership{Fellow,~IEEE} 
\thanks{\sloppy A. Eisenberg is with Bar-Ilan University, Israel, and OriginAI, Israel e-mail: \texttt{aviad.eisenberg@biu.ac.il}; S.E. Chazan is with OriginAI, Israel, email: \texttt{shlomi@originai.co}; and S. Gannot is with Bar-Ilan University, Israel, e-mail: \texttt{sharon.gannot@biu.ac.il}
}
}

\maketitle



\begin{abstract}
We present a multi-channel target speaker extraction (TSE) framework that is guided by relative spatial cues rather than explicit speaker-dependent references, such as clean enrollment utterances or precise directions of arrival (DOA). The proposed approach encodes the target speaker through discrete, interpretable relations between concurrent speakers—such as left/right or near/far—referred to as the Relative Position Cue (RPC). This relational conditioning effectively leverages spatial information while avoiding explicit DOA estimation, thereby enabling more robust operation under realistic conditions.
We evaluate the proposed method across multiple datasets encompassing diverse acoustic conditions\rev{, including directional and diffuse babble noise}, varying numbers of microphones, mixtures with different numbers of speakers, and dynamic scenarios involving moving speakers. Experimental results demonstrate that the RPC-based approach consistently achieves strong extraction performance and generalizes well across all tested configurations, establishing it as a practical and interpretable alternative for spatially informed speaker extraction. \rev{Audio examples are available at \url{https://aviad3189.github.io/RPC-extraction/}.} 
\end{abstract}



\section{INTRODUCTION}

\IEEEPARstart{S}{peaker} separation and extraction have been central challenges in speech processing for more than three decades. Both single-channel~\cite{wang2018supervised} and multichannel approaches have been extensively studied~\cite{gannot2017consolidated}. The recent adoption of deep learning has led to substantial progress in single-channel \ac{BSS}~\cite{subakan2021attention,wang2023tf,avenstrup2025sepmamba,lutati2023separate,zhao2024mossformer2,xu2024tiger}.

While the primary objective of \ac{BSS} is to recover all speakers present in a mixture, many practical applications require extracting only a single speaker of interest, a task commonly referred to as \ac{TSE}. Existing \ac{TSE} approaches typically condition the extraction model on explicit speaker-dependent references, such as clean enrollment utterances, or on geometric information, such as an estimated \ac{DOA}. However, these assumptions substantially limit their practical applicability. Enrollment recordings may be unavailable or acoustically mismatched to the target environment, while \ac{DOA} estimates are often unreliable in reverberant, noisy, or dynamically changing acoustic scenes. \rev{As discussed in the comprehensive overview in~\cite{zmolikova2023neural}, \ac{TSE} systems mainly differ in the type of \emph{clue} used to identify the target speaker, and each clue type entails a different trade-off between the information it carries and the effort required to obtain it. We next briefly review the main clue types.}

A common approach is to condition the extraction network on a speaker embedding computed from a reference utterance of the target speaker. The embedding captures speaker-specific characteristics and guides the network to selectively extract the desired speech from the mixture~\cite{zmolikova2023neural,eisenberg2022single,eisenberg2023two,ge2020spex+,hao2024x}. More recently, spatial conditioning has emerged as an attractive alternative to speaker embeddings. In particular, Zhang \emph{et al.}~\cite{zhang2025doa} presented a comprehensive comparison of spectral and spatial conditioning strategies, demonstrating that spatial information generally provides a more effective conditioning signal for target speaker extraction. Building on this observation, \cite{eisenberg2026spectral} proposed a systematic framework for adaptively leveraging spectral and spatial conditioning based on their inferred reliability.

A different paradigm is to perform blind multichannel speaker separation, recovering all speakers present in the mixture without requiring explicit conditioning. A prominent example is SpatialNet~\cite{quan2024spatialnet}, which effectively exploits multichannel spatial information to separate all speakers in the mixture. An extension of SpatialNet for dynamic acoustic scenarios was subsequently proposed in~\cite{quan2024multichannel}.

Although blind separation methods recover all speakers in the mixture, they leave the target selection problem unresolved. Consequently, even when the user's intention is specified spatially, an additional mechanism is required to identify the desired speaker among the separated outputs.

In multi-channel \ac{TSE}, a central challenge is determining which speaker to extract. Several strategies have been proposed, differing in the type of auxiliary information used to identify the target speaker.
Single-channel spectral information of the desired speaker has been used as a conditioning signal for multichannel separation models in~\cite{li2019direction,ge2022spex,han2021multi}. However, such approaches rely on prior familiarity with the target speaker, an assumption that is often unrealistic in real-world scenarios. In~\cite{taherian2024multi}, a joint diarization-separation was presented to overcome this limitation. First, an end-to-end diarization process is applied. Then, embeddings extracted from non-overlapping speech segments are used as reference guidance for the separation model, which estimates all speakers present in the mixture.   

The most natural form of spatial enrollment is the \ac{DOA} of the target speaker~\cite{tesch2023spatially,elminshawi2023beamformer}. Jing et al.~\cite{jing2025end} proposed an end-to-end framework that, given a \ac{DOA} and a specified beamwidth, extracts all speech signals within the corresponding directional sector while suppressing signals outside it. Pandey et al.~\cite{pandey2024all} presented a complementary approach in which the model receives frame-wise \ac{DOA} input, enabling rapid adaptation to moving speakers while maintaining low algorithmic latency, making it suitable for real-time applications. Another line of work uses audio samples captured at the target location as a spatial reference, either as raw recordings~\cite{martin2019multi} or as estimated \acp{RTF}~\cite{eisenberg2025end}.

It is worth noting that multimodal cues, such as text~\cite{dai2025inter,Ho2026Typing} and visual information~\cite{gu2020multi, xu2020neural, li2023audio, xu2023multi}, have also recently been used as reference enrollments to guide the extraction model. In this work, we focus on scenarios where only the audio modality is available.

In this paper, we introduce the \ac{RPC} as a spatially defined auxiliary signal to guide the proposed separation model. Concretely, RPC encodes the relative spatial relationship between two speakers, assuming only a minimal spatial separation. This relational specification is both robust and intuitive, since the target speaker can be identified by their position relative to the interferer. \rev{We stress that the \ac{RPC} specifies a \emph{relation between sources} (e.g., ``the target is to the right of the interferer''), rather than a \emph{region} of the room or a sector relative to the array. The boundary that separates the target from the interferer is therefore not fixed; it depends on the positions of all the concurrent speakers and must be inferred by the network from the mixture itself. This is the main difference from region-based clues~\cite{gu2024rezero,markovic2022implicit}, which fail when the target and the interferer reside in the same predefined region.} The proposed design exploits spatial cues---widely shown to benefit speaker extraction---while avoiding explicit \ac{DOA}/\ac{RTF} estimation and reducing reliance on potentially unreliable enrollment utterances. Moreover, because enrollment is expressed in relative rather than absolute terms, the approach naturally generalizes to moving-speaker settings, provided the speakers’ relative configuration remains unchanged.

\rev{In real conversations, overlapped speech occupies only a fraction of the total speaking time, and practical systems must also handle overlap-free segments and a varying number of active speakers. In this work, we focus on the overlapped segments, in which the target must be selected among several concurrent speakers. The scope of this assumption and the integration of the proposed method into a complete system are discussed in Sec.~\ref{sec:limitations}.}

Our main contributions are summarized as follows:
1) We introduce a novel spatial enrollment strategy based on relative position cues, enabling target speaker extraction without requiring explicit \ac{DOA} estimation or pre-recorded enrollment utterances;
2) We demonstrate that the proposed approach achieves strong performance across varying numbers of microphones and in mixtures containing multiple simultaneous speakers; and 3) We evaluate the proposed method under dynamic scenarios involving moving speakers, despite training exclusively on static mixtures, and demonstrate that it maintains robust extraction performance.

\section{Related works}
\rev{\label{sec:related}}

Beyond point-wise DOA conditioning, Gu et al. \cite{gu2024rezero} introduce a more general region-based conditioning strategy. Their model is trained to extract all speakers located within predefined spatial regions. Three types of spatial constraints are considered: angular sectors defined by DOA, circular regions defined by radius, and conical volumes that combine both DOA and radius. Directional features are derived from the similarity between inter-channel phase differences (IPD) and time-phase differences (TPD) within the query angle, while a learned embedding encodes the radius constraint. This formulation generalizes traditional DOA-based extraction by allowing the model to recover all sources in a queried region. However, an estimate of the DOA is still required.

Marković et al. proposed a similar approach in~\cite{markovic2022implicit}. Rather than focusing on precise point locations, their approach partitions the acoustic scene into two predefined spatial regions relative to the microphone array, such as left versus right or near versus far with respect to the array center. The model is then trained to extract all speakers located within the selected region. This provides an alternative to DOA-based methods. 

In contrast to these two works, the proposed approach does not rely on the absolute DOA of any speaker. Since the command is defined by the relative spatial configuration between speakers rather than their orientation with respect to the microphone array, explicit DOA estimation is unnecessary. Consequently, the proposed RPC enables separation even when all speakers are located on the same side of the microphone array, without requiring DOA information or prior knowledge of the speakers.

\rev{Region-based clues~\cite{gu2024rezero,markovic2022implicit,patterson2022distance} are indeed simpler than precise source locations, but they share a common property: the region is defined with respect to the microphone array by fixed (or user-specified) angular and/or distance boundaries. Such clues are well suited for extracting \emph{all} sources within a region of interest, but they cannot disambiguate two speakers that reside in the same region, e.g., two speakers that are both on the left of the array, or both closer than the distance threshold. The \ac{RPC}, in contrast, has no fixed boundary: ``left'' is defined with respect to the interfering speaker. The two approaches also address different tasks; ReZero~\cite{gu2024rezero} requires the absolute query angle and/or distance of the region and returns the sum of the sources inside it, whereas the \ac{RPC} returns a single speaker and requires no absolute geometric information. For this reason, we do not retrain ReZero as a baseline. Instead, in Sec.~\ref{sec:relation_vs_region}, we quantify how often \emph{any} fixed region boundary would fail to identify the target in our test data, and compare it with the behavior of the \ac{RPC}-based models.}

\section{Problem Formulation}

A scenario with $Q$ concurrent speakers, captured by $J$ microphones in a reverberant, noisy environment, is considered. The problem is analyzed in the \ac{STFT} domain, 
\begin{equation}
\label{eq:model}
\x = \sum_{q=1}^{Q}  \hq \cdot \sq   + \rev{\mathbf{v}(t,k)} ,
\end{equation}
where \( k \in \{0, \ldots, K-1\} \) and \( t \in \{0, \ldots, T-1\} \) denote the frequency and time-frame indices, respectively. Here, \( T \) and \( K \) represent the total number of time-frames and frequency bands. Let \( \sq \) represent the clean, anechoic speech signal of the \( q \)-th speaker,
where $\hq$ is a $J\times 1$ vector of the \acp{ATF} relating the $q$-th source and the microphones array\rev{, and $\mathbf{v}(t,k)$ is a $J\times 1$ vector of additive noise. This general formulation encompasses both directional and diffuse noise fields. For a single directional (point) noise source, $\mathbf{v}(t,k)=\hn \cdot n(t,k)$, where $\hn$ is the \acp{ATF} relating the noise and the microphones array, and $n(t,k)$ is the anechoic noise signal. In this case, the noise spatial covariance matrix is rank-1, and the mixture comprises $Q+1$ point sources. For diffuse noise, e.g., babble noise generated by many talkers distributed around the array, $\mathbf{v}(t,k)$ is the sum of many reverberant contributions arriving from all directions, and its spatial covariance matrix is full-rank, such that it cannot be cancelled by spatial nulling alone.} 
The reverberant target signal, as captured by the first microphone, is denoted  $\tilde{s}_d(t,k)$.
The goal is to extract the reverberant target speech signal, $\hat{\tilde{s}}_d(t,k)$, from the mixture $\mathbf{x}(t,k)$. This is achieved by utilizing spatial prior knowledge, denoted as the \emph{command} $C$, which indicates the desired speaker's position relative to the interferer\rev{(s). Throughout this work, we assume that all $Q$ speakers are concurrently active and that $Q$ is known}.

\section{Proposed Model}

In this section, we introduce a framework for speaker extraction that exploits relative spatial information and is compatible with a wide range of extraction architectures. We illustrate its application using two recently proposed \ac{TSE} schemes. The relative spatial information is defined with respect to a reference microphone array, providing cues about both the target speaker's angular position and distance from the array.
We first introduce the \acp{RPC} for two- and three-speaker settings. We then describe how the \acp{RPC} can be integrated into existing \ac{TSE} schemes. 

\subsection{\acf{RPC}: Two Speakers}

\begin{figure}
    \centering

\tikzset{every picture/.style={line width=0.75pt}} 

\begin{tikzpicture}[x=0.75pt,y=0.75pt,yscale=-1,xscale=1,scale=0.9,every node/.style={scale=0.9}]]

\draw  [fill={rgb, 255:red, 227; green, 216; blue, 216 }  ,fill opacity=1 ] (130,155.5) -- (490,155.5) -- (490,170) -- (130,170) -- cycle ;
\draw  [color={rgb, 255:red, 74; green, 144; blue, 226 }  ,draw opacity=1 ][fill={rgb, 255:red, 74; green, 144; blue, 226 }  ,fill opacity=1 ] (185.5,118.25) .. controls (185.5,115.35) and (187.85,113) .. (190.75,113) .. controls (193.65,113) and (196,115.35) .. (196,118.25) .. controls (196,121.15) and (193.65,123.5) .. (190.75,123.5) .. controls (187.85,123.5) and (185.5,121.15) .. (185.5,118.25) -- cycle ;
\draw  [color={rgb, 255:red, 208; green, 2; blue, 27 }  ,draw opacity=1 ][fill={rgb, 255:red, 208; green, 2; blue, 27 }  ,fill opacity=1 ] (185.5,96.25) .. controls (185.5,93.35) and (187.85,91) .. (190.75,91) .. controls (193.65,91) and (196,93.35) .. (196,96.25) .. controls (196,99.15) and (193.65,101.5) .. (190.75,101.5) .. controls (187.85,101.5) and (185.5,99.15) .. (185.5,96.25) -- cycle ;
\draw   (130,80) -- (250,80) -- (250,170) -- (130,170) -- cycle ;
\draw (175.07,144.86) node  {\includegraphics[width=4.39pt,height=7.29pt]{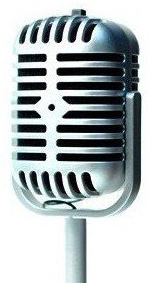}};
\draw (185.07,144.86) node  {\includegraphics[width=4.39pt,height=7.29pt]{images/mic.png}};
\draw (195.07,144.86) node  {\includegraphics[width=4.39pt,height=7.29pt]{images/mic.png}};
\draw (205.07,144.86) node  {\includegraphics[width=4.39pt,height=7.29pt]{images/mic.png}};

\draw   (250,80) -- (370,80) -- (370,170) -- (250,170) -- cycle ;
\draw (295.07,144.86) node  {\includegraphics[width=4.39pt,height=7.29pt]{images/mic.png}};
\draw (305.07,144.86) node  {\includegraphics[width=4.39pt,height=7.29pt]{images/mic.png}};
\draw (315.07,144.86) node  {\includegraphics[width=4.39pt,height=7.29pt]{images/mic.png}};
\draw (325.07,144.86) node  {\includegraphics[width=4.39pt,height=7.29pt]{images/mic.png}};

\draw  [color={rgb, 255:red, 208; green, 2; blue, 27 }  ,draw opacity=1 ][fill={rgb, 255:red, 208; green, 2; blue, 27 }  ,fill opacity=1 ] (333,111.25) .. controls (333,108.35) and (335.35,106) .. (338.25,106) .. controls (341.15,106) and (343.5,108.35) .. (343.5,111.25) .. controls (343.5,114.15) and (341.15,116.5) .. (338.25,116.5) .. controls (335.35,116.5) and (333,114.15) .. (333,111.25) -- cycle ;
\draw  [color={rgb, 255:red, 74; green, 144; blue, 226 }  ,draw opacity=1 ][fill={rgb, 255:red, 74; green, 144; blue, 226 }  ,fill opacity=1 ] (274,110.75) .. controls (274,107.85) and (276.35,105.5) .. (279.25,105.5) .. controls (282.15,105.5) and (284.5,107.85) .. (284.5,110.75) .. controls (284.5,113.65) and (282.15,116) .. (279.25,116) .. controls (276.35,116) and (274,113.65) .. (274,110.75) -- cycle ;
\draw  [color={rgb, 255:red, 74; green, 144; blue, 226 }  ,draw opacity=1 ][fill={rgb, 255:red, 74; green, 144; blue, 226 }  ,fill opacity=1 ] (400,126) .. controls (400,122.69) and (402.69,120) .. (406,120) .. controls (409.31,120) and (412,122.69) .. (412,126) .. controls (412,129.31) and (409.31,132) .. (406,132) .. controls (402.69,132) and (400,129.31) .. (400,126) -- cycle ;
\draw  [color={rgb, 255:red, 208; green, 2; blue, 27 }  ,draw opacity=1 ][fill={rgb, 255:red, 208; green, 2; blue, 27 }  ,fill opacity=1 ] (450,95.25) .. controls (450,92.35) and (452.35,90) .. (455.25,90) .. controls (458.15,90) and (460.5,92.35) .. (460.5,95.25) .. controls (460.5,98.15) and (458.15,100.5) .. (455.25,100.5) .. controls (452.35,100.5) and (450,98.15) .. (450,95.25) -- cycle ;
\draw   (370,80) -- (490,80) -- (490,170) -- (370,170) -- cycle ;
\draw (415.07,144.86) node  {\includegraphics[width=4.39pt,height=7.29pt]{images/mic.png}};
\draw (425.07,144.86) node  {\includegraphics[width=4.39pt,height=7.29pt]{images/mic.png}};
\draw (435.07,144.86) node  {\includegraphics[width=4.39pt,height=7.29pt]{images/mic.png}};
\draw (445.07,144.86) node  {\includegraphics[width=4.39pt,height=7.29pt]{images/mic.png}};

\draw (478.9,99.1) node  {\includegraphics[width=8.85pt,height=8.85pt]{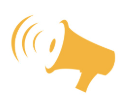}};
\draw (358.9,99.9) node  {\includegraphics[width=8.85pt,height=8.85pt]{images/megaphone2.png}};
\draw (238.1,99.1) node  {\includegraphics[width=8.85pt,height=8.85pt]{images/megaphone2.png}};

\draw (423,154) node [anchor=north west][inner sep=0.75pt]  [font=\small] [align=left] {{\fontfamily{ptm}\selectfont {\footnotesize (c)}}};
\draw (305,154) node [anchor=north west][inner sep=0.75pt]  [font=\small] [align=left] {{\fontfamily{ptm}\selectfont {\footnotesize (b)}}};
\draw (184,154) node [anchor=north west][inner sep=0.75pt]  [font=\small] [align=left] {{\fontfamily{ptm}\selectfont {\footnotesize (a)}}};

\end{tikzpicture}

    \caption{In the two-speaker scenario, we consider three distinct spatial configurations of the speakers relative to the center of the array:
(a) Both speakers share the same direction of arrival but are located at different distances from the center of the array;
(b) Both speakers are equidistant from the center but differ in their directions of arrival;
and (c) The speakers differ in both their distances from the array center and their directions of arrival.
}

\label{fig:scenarios}
\end{figure}

 Speakers may differ in their lateral position (left–right) and in their distance from the array. This formulation is consistent with human perception: when two speakers are positioned in front of a listener, they can be readily distinguished by spatial cues such as horizontal separation and depth.

Assuming a two-concurrent-speaker scenario ($Q = 2$), the spatial relationship yields eight \acp{RPC}, which can be used to identify the target speaker, as summarized in Table~\ref{table:speaker_positions}.
\begin{table}[h!]
\centering
\caption{{Relative speaker positions and orientations for the two-source configuration.}}
\label{table:speaker_positions}
\resizebox{\columnwidth}{!}{%
\renewcommand{\arraystretch}{0.85}
\begin{tabular}{cl}
\toprule
\textbf{\#} & \textbf{\acp{RPC} for two speakers scenario:} \\ \midrule
\rev{1} & \rev{The target speaker is farther from the array than the interferer} \\
\rev{2} & \rev{The target speaker is closer to the array than the interferer} \\
\rev{3} & \rev{The target speaker is to the right of the interferer} \\
\rev{4} & \rev{The target speaker is to the left of the interferer} \\
\rev{5} & \rev{The target speaker is farther than and to the right of the interferer} \\
\rev{6} & \rev{The target speaker is closer than and to the left of the interferer} \\
\rev{7} & \rev{The target speaker is closer than and to the right of the interferer} \\
\rev{8} & \rev{The target speaker is farther than and to the left of the interferer} \\ \bottomrule
\end{tabular}
}
\end{table}
These eight options can be categorized into three groups:
\rev{(1) \textit{vertical-only information}, which reflects distance-related cues (\#1--2, Fig.~\ref{fig:scenarios}(a));
(2) \textit{horizontal-only information}, which corresponds to \ac{DOA} cues (\#3--4, Fig.~\ref{fig:scenarios}(b)); and}
(3) \textit{combined horizontal and vertical information}, incorporating both directional and distance cues\rev{ (\#5--8, Fig.~\ref{fig:scenarios}(c) and its mirrored and inverted versions)}.
\rev{Here, ``left'' and ``right'' are defined from the viewpoint of the array.}

A model can be trained using either the given \ac{DOA} and distance, a textual prompt indicating the target speaker, or a prefix integer corresponding to each possible configuration. Given that the number of distinct cases is finite (eight in total), we opted for an index-based prompt rather than textual descriptions, as the latter were deemed unnecessarily complex for this setting.

To characterize a given scenario within the proposed categories, we define two key spatial metrics. The gap distance between the speakers is expressed as:
\begin{equation}
    \psi = | R_1-R_2|,
\end{equation}
and the \ac{DOA} distance is given by:
\begin{equation}
    \phi = | \theta_1-\theta_2|,
\end{equation}
where $R_i$ denotes the distance of speaker $i$ from the center of the microphone array, and $\theta_i$ represents the \ac{DOA} angle of speaker $i$.

The speakers are considered to be radially separated if the gap distance satisfies $\psi > \psi_{min}$, indicating that they are located sufficiently far from each other in terms of radial distance. Furthermore, a relative \ac{DOA} separation is identified when it exceeds a threshold, i.e, $\phi > \phi_{min}$, implying a significant angular displacement between the speakers.
\rev{Throughout this work, we use $\psi_{min}=0.5$~m and $\phi_{min}=10^\circ$.}

\rev{\textit{How the two metrics are used.} The metrics $\psi$ and $\phi$ are used only to \emph{assign} the command to a training or test mixture; they are not given to the network. Exactly one command is assigned to each speaker in each mixture: if both $\psi>\psi_{min}$ and $\phi>\phi_{min}$, a combined cue (\#5--8) is used; if only $\psi>\psi_{min}$, a distance-only cue (\#1--2) is used; and if only $\phi>\phi_{min}$, a direction-only cue (\#3--4) is used. Note that the noise source is not treated as a source in the definition of the cue. The command is fed to the network as an integer index (Sec.~\ref{sec:integration}); the textual descriptions in Table~\ref{table:speaker_positions} are given for readability only.}

\rev{\textit{Limitations imposed by the thresholds.} Configurations with $\psi\le\psi_{min}$ \emph{and} $\phi\le\phi_{min}$ are not covered by the proposed cues, since the speakers cannot be reliably told apart by their relative position. Two speakers can share the same radius $R$ (e.g., when positioned symmetrically with respect to the array broadside); such configurations, in which $\psi$ is arbitrarily small, do appear in our data, and in this case only a direction-only cue is used. Similarly, sources that share the same \ac{DOA} are distinguished only by a distance-only cue. Finally, since a \ac{ULA} cannot resolve front--back ambiguity, all sources are placed in the half-plane in front of the array, i.e., $\theta\in[0^\circ,180^\circ]$.}

\subsection{\acf{RPC}: Three Speakers}
\rev{\label{sec:general_q}}

As the number of speakers increases, the complexity of the \ac{RPC} also grows, since the spatial relationships among multiple speakers are inherently more intricate than in the case of only two speakers. In this scenario, we define the desired speaker as being isolated from the two interfering speakers either by radial distance, by \ac{DOA}, or both. The two interfering speakers may be positioned very close to one another--potentially nearly indistinguishable in spatial terms--as long as the desired speaker is well separated from them in at least one of these two domains.

The spatial metrics are redefined as follows. The radial distance between speakers is expressed as:
\begin{equation}
    \psi = \min \{| R_d-R_{i_1}|, | R_d-R_{i_2}| \}
\end{equation}
and the DOA distance is given by: 
\begin{equation}
    \phi = \min \{| \theta_d-\theta_{i_1}|,  |\theta_d-\theta_{i_2}| \}
\end{equation}
where $R_d$ denotes the distance of the target speaker from the center of the microphone array, while $R_{i_1}$ and $R_{i_2}$ represent the distances of the interfering speakers, which are selected randomly. Similarly, $\theta_d$ corresponds to the \ac{DOA} of the desired speaker, and $\theta_{i_1}$ and $\theta_{i_2}$ denote the \ac{DOA} of the interfering speakers, which are also randomly assigned in correspondence with their respective radial distances.

As before, the same conditions must be satisfied for the desired speaker to be considered spatially separated from the interfering sources.
In this setup, the {command} strategy is slightly modified to define the target speaker by specifying one of the nine spatial positions shown in Table \ref{table:speaker_positions_3}.

\begin{table}[h!]
\centering
\caption{{Relative speaker positions and orientations for the three-source configuration.}}
\label{table:speaker_positions_3}
\resizebox{\columnwidth}{!}{%
\renewcommand{\arraystretch}{0.85}
\begin{tabular}{cl}
\toprule
\textbf{\#} & \textbf{Three speakers \acp{RPC}:} \\ \midrule
\rev{1} & \rev{The target is the farthest and the left-most speaker} \\
\rev{2} & \rev{The target is the farthest speaker, between the other two} \\
\rev{3} & \rev{The target is the farthest and the right-most speaker} \\
\rev{4} & \rev{The target is the middle (in distance) and the left-most speaker} \\
\rev{5} & \rev{The target is the middle speaker in both distance and direction} \\
\rev{6} & \rev{The target is the middle (in distance) and the right-most speaker} \\
\rev{7} & \rev{The target is the closest and the left-most speaker} \\
\rev{8} & \rev{The target is the closest speaker, between the other two} \\
\rev{9} & \rev{The target is the closest and the right-most speaker} \\ \bottomrule
\end{tabular}
}
\end{table}

\begin{figure}
\centering

\tikzset{every picture/.style={line width=0.75pt}} 

\begin{tikzpicture}[x=0.75pt,y=0.75pt,yscale=-1,xscale=1,scale=0.9,every node/.style={scale=0.9}]]

\draw  [fill={rgb, 255:red, 227; green, 216; blue, 216 }  ,fill opacity=1 ] (130,265.5) -- (490,265.5) -- (490,280) -- (130,280) -- cycle ;
\draw  [color={rgb, 255:red, 74; green, 144; blue, 226 }  ,draw opacity=1 ][fill={rgb, 255:red, 74; green, 144; blue, 226 }  ,fill opacity=1 ] (199,226.75) .. controls (199,223.85) and (201.35,221.5) .. (204.25,221.5) .. controls (207.15,221.5) and (209.5,223.85) .. (209.5,226.75) .. controls (209.5,229.65) and (207.15,232) .. (204.25,232) .. controls (201.35,232) and (199,229.65) .. (199,226.75) -- cycle ;
\draw  [color={rgb, 255:red, 208; green, 2; blue, 27 }  ,draw opacity=1 ][fill={rgb, 255:red, 208; green, 2; blue, 27 }  ,fill opacity=1 ] (185.5,206.25) .. controls (185.5,203.35) and (187.85,201) .. (190.75,201) .. controls (193.65,201) and (196,203.35) .. (196,206.25) .. controls (196,209.15) and (193.65,211.5) .. (190.75,211.5) .. controls (187.85,211.5) and (185.5,209.15) .. (185.5,206.25) -- cycle ;
\draw   (130,190) -- (250,190) -- (250,280) -- (130,280) -- cycle ;
\draw (175.07,254.86) node  {\includegraphics[width=4.39pt,height=7.29pt]{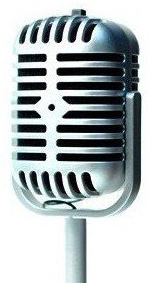}};
\draw (185.07,254.86) node  {\includegraphics[width=4.39pt,height=7.29pt]{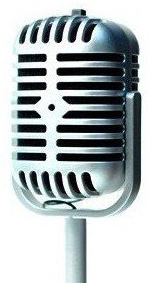}};
\draw (195.07,254.86) node  {\includegraphics[width=4.39pt,height=7.29pt]{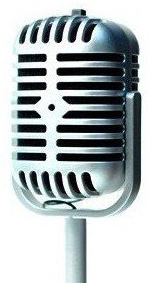}};
\draw (205.07,254.86) node  {\includegraphics[width=4.39pt,height=7.29pt]{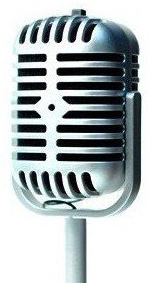}};

\draw  [color={rgb, 255:red, 74; green, 144; blue, 226 }  ,draw opacity=1 ][fill={rgb, 255:red, 74; green, 144; blue, 226 }  ,fill opacity=1 ] (164.5,228.25) .. controls (164.5,225.35) and (166.85,223) .. (169.75,223) .. controls (172.65,223) and (175,225.35) .. (175,228.25) .. controls (175,231.15) and (172.65,233.5) .. (169.75,233.5) .. controls (166.85,233.5) and (164.5,231.15) .. (164.5,228.25) -- cycle ;
\draw  [color={rgb, 255:red, 237; green, 50; blue, 102 }  ,draw opacity=1 ][fill={rgb, 255:red, 237; green, 50; blue, 102 }  ,fill opacity=0.93 ] (217.5,199.75) .. controls (217.5,196.85) and (219.85,194.5) .. (222.75,194.5) .. controls (225.65,194.5) and (228,196.85) .. (228,199.75) .. controls (228,202.65) and (225.65,205) .. (222.75,205) .. controls (219.85,205) and (217.5,202.65) .. (217.5,199.75) -- cycle ;
\draw  [color={rgb, 255:red, 239; green, 134; blue, 146 }  ,draw opacity=1 ][fill={rgb, 255:red, 239; green, 134; blue, 146 }  ,fill opacity=1 ] (143.5,203.75) .. controls (143.5,200.85) and (145.85,198.5) .. (148.75,198.5) .. controls (151.65,198.5) and (154,200.85) .. (154,203.75) .. controls (154,206.65) and (151.65,209) .. (148.75,209) .. controls (145.85,209) and (143.5,206.65) .. (143.5,203.75) -- cycle ;
\draw   (250,190) -- (370,190) -- (370,280) -- (250,280) -- cycle ;
\draw (295.07,254.86) node  {\includegraphics[width=4.39pt,height=7.29pt]{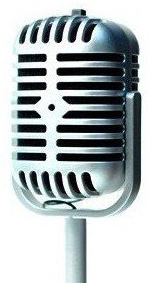}};
\draw (305.07,254.86) node  {\includegraphics[width=4.39pt,height=7.29pt]{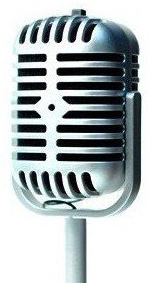}};
\draw (315.07,254.86) node  {\includegraphics[width=4.39pt,height=7.29pt]{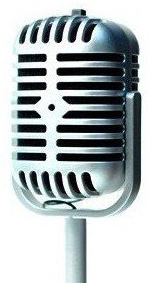}};
\draw (325.07,254.86) node  {\includegraphics[width=4.39pt,height=7.29pt]{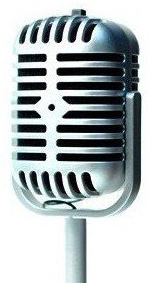}};

\draw  [color={rgb, 255:red, 237; green, 50; blue, 102 }  ,draw opacity=1 ][fill={rgb, 255:red, 237; green, 50; blue, 102 }  ,fill opacity=0.93 ] (338,217.75) .. controls (338,214.85) and (340.35,212.5) .. (343.25,212.5) .. controls (346.15,212.5) and (348.5,214.85) .. (348.5,217.75) .. controls (348.5,220.65) and (346.15,223) .. (343.25,223) .. controls (340.35,223) and (338,220.65) .. (338,217.75) -- cycle ;
\draw  [color={rgb, 255:red, 74; green, 144; blue, 226 }  ,draw opacity=1 ][fill={rgb, 255:red, 74; green, 144; blue, 226 }  ,fill opacity=1 ] (286.5,231.75) .. controls (286.5,228.85) and (288.85,226.5) .. (291.75,226.5) .. controls (294.65,226.5) and (297,228.85) .. (297,231.75) .. controls (297,234.65) and (294.65,237) .. (291.75,237) .. controls (288.85,237) and (286.5,234.65) .. (286.5,231.75) -- cycle ;
\draw  [color={rgb, 255:red, 74; green, 144; blue, 226 }  ,draw opacity=1 ][fill={rgb, 255:red, 74; green, 144; blue, 226 }  ,fill opacity=1 ] (324.5,200.25) .. controls (324.5,197.35) and (326.85,195) .. (329.75,195) .. controls (332.65,195) and (335,197.35) .. (335,200.25) .. controls (335,203.15) and (332.65,205.5) .. (329.75,205.5) .. controls (326.85,205.5) and (324.5,203.15) .. (324.5,200.25) -- cycle ;
\draw  [color={rgb, 255:red, 208; green, 2; blue, 27 }  ,draw opacity=1 ][fill={rgb, 255:red, 208; green, 2; blue, 27 }  ,fill opacity=1 ] (305,217.75) .. controls (305,214.85) and (307.35,212.5) .. (310.25,212.5) .. controls (313.15,212.5) and (315.5,214.85) .. (315.5,217.75) .. controls (315.5,220.65) and (313.15,223) .. (310.25,223) .. controls (307.35,223) and (305,220.65) .. (305,217.75) -- cycle ;
\draw  [color={rgb, 255:red, 239; green, 134; blue, 146 }  ,draw opacity=1 ][fill={rgb, 255:red, 239; green, 134; blue, 146 }  ,fill opacity=1 ] (271,213.75) .. controls (271,210.85) and (273.35,208.5) .. (276.25,208.5) .. controls (279.15,208.5) and (281.5,210.85) .. (281.5,213.75) .. controls (281.5,216.65) and (279.15,219) .. (276.25,219) .. controls (273.35,219) and (271,216.65) .. (271,213.75) -- cycle ;

\draw  [color={rgb, 255:red, 74; green, 144; blue, 226 }  ,draw opacity=1 ][fill={rgb, 255:red, 74; green, 144; blue, 226 }  ,fill opacity=1 ] (448.5,209) .. controls (448.5,205.69) and (451.19,203) .. (454.5,203) .. controls (457.81,203) and (460.5,205.69) .. (460.5,209) .. controls (460.5,212.31) and (457.81,215) .. (454.5,215) .. controls (451.19,215) and (448.5,212.31) .. (448.5,209) -- cycle ;
\draw  [color={rgb, 255:red, 208; green, 2; blue, 27 }  ,draw opacity=1 ][fill={rgb, 255:red, 208; green, 2; blue, 27 }  ,fill opacity=1 ] (432,227.25) .. controls (432,224.35) and (434.35,222) .. (437.25,222) .. controls (440.15,222) and (442.5,224.35) .. (442.5,227.25) .. controls (442.5,230.15) and (440.15,232.5) .. (437.25,232.5) .. controls (434.35,232.5) and (432,230.15) .. (432,227.25) -- cycle ;
\draw   (370,190) -- (490,190) -- (490,280) -- (370,280) -- cycle ;
\draw (415.07,254.86) node  {\includegraphics[width=4.39pt,height=7.29pt]{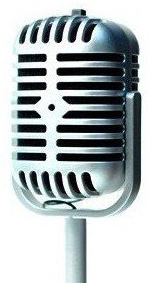}};
\draw (425.07,254.86) node  {\includegraphics[width=4.39pt,height=7.29pt]{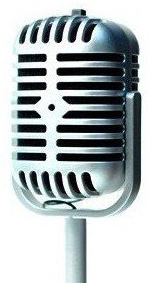}};
\draw (435.07,254.86) node  {\includegraphics[width=4.39pt,height=7.29pt]{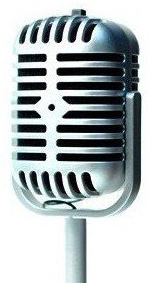}};
\draw (445.07,254.86) node  {\includegraphics[width=4.39pt,height=7.29pt]{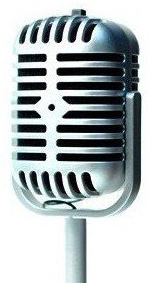}};

\draw  [color={rgb, 255:red, 74; green, 144; blue, 226 }  ,draw opacity=1 ][fill={rgb, 255:red, 74; green, 144; blue, 226 }  ,fill opacity=1 ] (414,202.25) .. controls (414,199.35) and (416.35,197) .. (419.25,197) .. controls (422.15,197) and (424.5,199.35) .. (424.5,202.25) .. controls (424.5,205.15) and (422.15,207.5) .. (419.25,207.5) .. controls (416.35,207.5) and (414,205.15) .. (414,202.25) -- cycle ;
\draw  [color={rgb, 255:red, 237; green, 50; blue, 102 }  ,draw opacity=1 ][fill={rgb, 255:red, 237; green, 50; blue, 102 }  ,fill opacity=0.93 ] (466,229.25) .. controls (466,226.35) and (468.35,224) .. (471.25,224) .. controls (474.15,224) and (476.5,226.35) .. (476.5,229.25) .. controls (476.5,232.15) and (474.15,234.5) .. (471.25,234.5) .. controls (468.35,234.5) and (466,232.15) .. (466,229.25) -- cycle ;
\draw  [color={rgb, 255:red, 239; green, 134; blue, 146 }  ,draw opacity=1 ][fill={rgb, 255:red, 239; green, 134; blue, 146 }  ,fill opacity=1 ] (393.5,228.75) .. controls (393.5,225.85) and (395.85,223.5) .. (398.75,223.5) .. controls (401.65,223.5) and (404,225.85) .. (404,228.75) .. controls (404,231.65) and (401.65,234) .. (398.75,234) .. controls (395.85,234) and (393.5,231.65) .. (393.5,228.75) -- cycle ;

\draw (238.9,210.1) node  {\includegraphics[width=8.85pt,height=8.85pt]{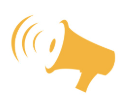}};
\draw (479.3,210.3) node  {\includegraphics[width=8.85pt,height=8.85pt]{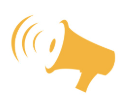}};
\draw (360.1,209.1) node  {\includegraphics[width=8.85pt,height=8.85pt]{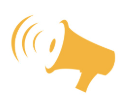}};

\draw (183,264) node [anchor=north west][inner sep=0.75pt]  [font=\small] [align=left] {{\fontfamily{ptm}\selectfont {\footnotesize (a)}}};
\draw (304.14,263.71) node [anchor=north west][inner sep=0.75pt]  [font=\small] [align=left] {{\fontfamily{ptm}\selectfont {\footnotesize (b)}}};
\draw (424,263.71) node [anchor=north west][inner sep=0.75pt]  [font=\small] [align=left] {{\fontfamily{ptm}\selectfont {\footnotesize (c)}}};

\end{tikzpicture}

\caption{An illustration of the nine possible \ac{RPC}s for the three-speaker scenario.
The target speaker {(in red)} is defined by its relative position with respect to the two interfering sources, characterized by differences in radial distance and/or \ac{DOA}: (a) The far speaker positioned on the \textcolor[rgb]{0.94,0.53,0.57}{left} / \textcolor[rgb]{0.82,0.01,0.11}{center} /  \textcolor[rgb]{0.93,0.2,0.4}{right}; (b) The middle speaker positioned on the  \textcolor[rgb]{0.94,0.53,0.57}{left} / \textcolor[rgb]{0.82,0.01,0.11}{center} /  \textcolor[rgb]{0.93,0.2,0.4}{right}; (c) The near speaker positioned on the \textcolor[rgb]{0.94,0.53,0.57}{ left} / \textcolor[rgb]{0.82,0.01,0.11}{center} /  \textcolor[rgb]{0.93,0.2,0.4}{right}.}
\label{fig:three_speakers_scenarios}
\end{figure}

These nine categories comprehensively cover the possible configurations for isolating the target speaker. An illustration of these scenarios can be found in Fig.~\ref{fig:three_speakers_scenarios}. {In both the two- and three-speaker configurations, the user must accurately identify the desired speaker.} 

\vspace{-10pt}

\subsection{Incorporating \ac{RPC} Information into \ac{TSE} Schemes}
\rev{\label{sec:integration}}
The proposed \ac{RPC} framework is architecture-agnostic and can be integrated into a wide range of target speech extraction and speech separation systems. The key idea is to represent each \ac{RPC} command using a learnable embedding and inject this representation into the network to guide the extraction process toward the desired speaker. To demonstrate the generality and effectiveness of the proposed approach, we incorporated the \ac{RPC} framework into two previously published multichannel architectures: (i) our U-Net-based speaker extraction model \cite{eisenberg2025end}, and (ii) SpatialNet \cite{quan2024spatialnet}. The following subsections detail the modifications required to integrate \ac{RPC} into each architecture.

\subsubsection{Integration with a U-Net-Based Speaker Extraction Model}
We first incorporated the proposed framework into our previously published multichannel speaker extraction model \cite{eisenberg2025end}. The model comprises a U-Net architecture with a self-attention mechanism in the bottleneck. The encoder consists of several Conv2D layers, each followed by batch normalization to improve training stability and generalization. A \ac{PReLU} activation function is applied after each layer. Subsequently, the channel and frequency dimensions are combined, and a fully connected layer is used to reduce the resulting dimensionality. Finally, a self-attention layer is applied to model long-range dependencies in the latent representation.

The decoder employs transposed convolution layers whose dimensions match those of the corresponding encoder layers. This design enables skip connections between encoder and decoder stages, facilitating information flow and preserving fine-grained details throughout the reconstruction process.

To incorporate \ac{RPC} information, each command is assigned a unique integer identifier. A learnable lookup table is jointly trained to generate a dedicated embedding for every command. The embedding corresponding to the selected command is then used to condition the mixture representation through \ac{FiLM} \cite{perez2017film}. Formally,
\begin{equation}
    \text{FiLM}(\text{emb}_x,\text{emb}_c)
    =
    \text{emb}_x \cdot \gamma(\text{emb}_c)
    +
    \beta(\text{emb}_c),
\end{equation}
where $\text{emb}_x$ denotes the mixture embedding, $\text{emb}_c$ denotes the command embedding, and $\gamma(\cdot)$ and $\beta(\cdot)$ are learned linear transformations. An illustration of the resulting architecture is presented in Fig.~\ref{fig:subfig_b}.

\subsubsection{Integration with SpatialNet}
We further incorporated the proposed \ac{RPC} framework into SpatialNet \cite{quan2024spatialnet}, a multichannel speech separation architecture composed of alternating narrow-band and cross-band processing blocks. The narrow-band blocks model temporal-spatial dependencies within each frequency band using a self-attention module followed by a convolution-augmented feed-forward module. The cross-band blocks operate across frequency bands and capture inter-band dependencies.

\begin{figure}[htbp]
    \centering

    \begin{subfigure}{0.48\textwidth}
        \centering
        \begin{tikzpicture}[x=0.75pt,y=0.75pt,yscale=-1,xscale=1,scale=0.9]

\draw  [fill={rgb, 255:red, 235; green, 199; blue, 204 }  ,fill opacity=1 ] (225.7,380.31) -- (269.95,405.97) -- (269.92,427.38) -- (225.6,452.92) -- cycle ;

\draw   (275.89,408.88) .. controls (275.89,406.68) and (277.68,404.88) .. (279.89,404.88) -- (304.89,404.88) .. controls (307.1,404.88) and (308.89,406.68) .. (308.89,408.88) -- (308.89,420.88) .. controls (308.89,423.09) and (307.1,424.88) .. (304.89,424.88) -- (279.89,424.88) .. controls (277.68,424.88) and (275.89,423.09) .. (275.89,420.88) -- cycle ;

\draw  [fill={rgb, 255:red, 178; green, 200; blue, 225 }  ,fill opacity=1 ] (314.89,388.38) .. controls (314.89,387.28) and (315.78,386.38) .. (316.89,386.38) -- (322.89,386.38) .. controls (323.99,386.38) and (324.89,387.28) .. (324.89,388.38) -- (324.89,444.38) .. controls (324.89,445.49) and (323.99,446.38) .. (322.89,446.38) -- (316.89,446.38) .. controls (315.78,446.38) and (314.89,445.49) .. (314.89,444.38) -- cycle ;

\draw  [fill={rgb, 255:red, 235; green, 199; blue, 204 }  ,fill opacity=1 ] (377.3,452.92) -- (332.97,427.4) -- (332.93,405.99) -- (377.17,380.31) -- cycle ;
\draw    (378.6,416) -- (405.6,416) ;
\draw [shift={(407.6,416)}, rotate = 180] [color={rgb, 255:red, 0; green, 0; blue, 0 }  ][line width=0.75]    (10.93,-3.29) .. controls (6.95,-1.4) and (3.31,-0.3) .. (0,0) .. controls (3.31,0.3) and (6.95,1.4) .. (10.93,3.29)   ;
\draw   (417.6,383.71) -- (476.63,383.71) -- (476.63,442.33) -- (417.6,442.33) -- cycle ;
\draw (447.12,413.02) node  {\includegraphics[width=44.27pt,height=43.96pt]{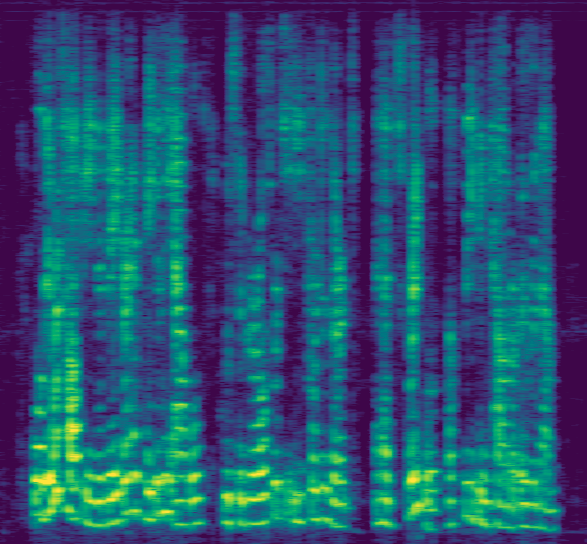}};

\draw   (140.1,373.31) -- (199.13,373.31) -- (199.13,431.93) -- (140.1,431.93) -- cycle ;
\draw (169.62,402.62) node  {\includegraphics[width=44.27pt,height=43.96pt]{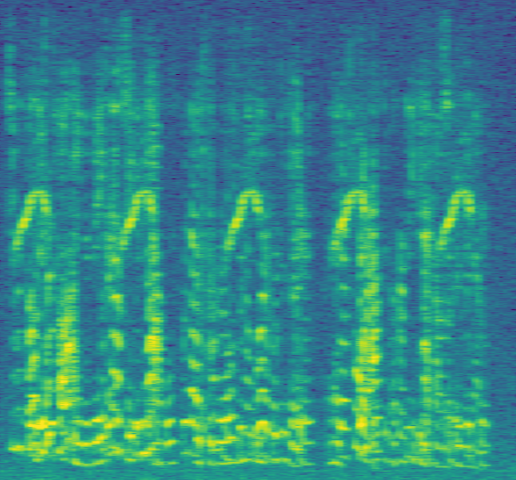}};

\draw   (143.1,376.71) -- (202.13,376.71) -- (202.13,435.33) -- (143.1,435.33) -- cycle ;
\draw (172.62,406.02) node  {\includegraphics[width=44.27pt,height=43.96pt]{images/mix.png}};

\draw   (147.1,380.71) -- (206.13,380.71) -- (206.13,439.33) -- (147.1,439.33) -- cycle ;
\draw (176.62,410.02) node  {\includegraphics[width=44.27pt,height=43.96pt]{images/mix.png}};

\draw   (150.1,384.11) -- (209.13,384.11) -- (209.13,442.73) -- (150.1,442.73) -- cycle ;
\draw (179.62,413.42) node  {\includegraphics[width=44.27pt,height=43.96pt]{images/mix.png}};

\draw   (203.73,328.79) -- (138.53,328.88) -- (138.52,319.21) -- (203.72,319.12) -- cycle ;
\draw   (203.74,338.45) -- (138.55,338.55) -- (138.53,328.88) -- (203.73,328.79) -- cycle ;
\draw   (203.76,348.12) -- (138.56,348.21) -- (138.55,338.54) -- (203.74,338.45) -- cycle ;
\draw  [fill={rgb, 255:red, 245; green, 166; blue, 35 }  ,fill opacity=1 ] (203.77,357.79) -- (138.57,357.88) -- (138.56,348.21) -- (203.76,348.12) -- cycle ;
\draw   (203.78,367.45) -- (138.59,367.54) -- (138.57,357.88) -- (203.77,357.79) -- cycle ;
\draw   (203.72,319.12) -- (138.52,319.21) -- (138.5,309.54) -- (203.7,309.45) -- cycle ;

\draw    (206.14,338.43) -- (293,338.43) -- (293,393.25) ;
\draw [shift={(293,395.25)}, rotate = 270] [color={rgb, 255:red, 0; green, 0; blue, 0 }  ][line width=0.75]    (10.93,-3.29) .. controls (6.95,-1.4) and (3.31,-0.3) .. (0,0) .. controls (3.31,0.3) and (6.95,1.4) .. (10.93,3.29)   ;

\draw (337.93,408.99) node [anchor=north west][inner sep=0.75pt]   [align=left] {{\scriptsize {\fontfamily{ptm}\selectfont Decoder}}};
\draw (306.31,345.87) node [anchor=north west][inner sep=0.75pt]  [font=\scriptsize,rotate=-90.83]  {$E_{emb}$};
\draw (142.67,299) node [anchor=north west][inner sep=0.75pt]   [align=left] {{\tiny {\fontfamily{pcr}\selectfont lookup table}}};
\draw (184.8,447.4) node [anchor=north west][inner sep=0.75pt]   [align=left] {{\tiny [T,K,2J]}};
\draw (456.1,447) node [anchor=north west][inner sep=0.75pt]   [align=left] {{\tiny [T,K,2]}};
\draw (312.75,444.39) node [anchor=north west][inner sep=0.75pt]  [font=\scriptsize,rotate=-270.2] [align=left] {{\scriptsize self-attention}};
\draw (281.39,409.38) node [anchor=north west][inner sep=0.75pt]   [align=left] {{\tiny {\fontfamily{pcr}\selectfont FiLM}}};
\draw (228.89,408.98) node [anchor=north west][inner sep=0.75pt]   [align=left] {{\fontfamily{ptm}\selectfont {\scriptsize Encoder}}};

\end{tikzpicture}
        \caption{Unet-based architecture}
        \label{fig:subfig_b}
    \end{subfigure}

    \begin{subfigure}{0.48\textwidth}
        \centering
     \begin{tikzpicture}[x=0.75pt,y=0.75pt,yscale=-1,xscale=1,scale=0.9]

\draw    (121.6,101.11) -- (180.63,101.11) -- (180.63,159.73) -- (121.6,159.73) -- cycle ;
\draw (151.12,130.42) node  {\includegraphics[width=44.27pt,height=43.96pt]{images/mix.png}};
\draw    (124.6,104.51) -- (183.63,104.51) -- (183.63,163.13) -- (124.6,163.13) -- cycle ;
\draw (154.12,133.82) node  {\includegraphics[width=44.27pt,height=43.96pt]{images/mix.png}};
\draw    (128.6,108.51) -- (187.63,108.51) -- (187.63,167.13) -- (128.6,167.13) -- cycle ;
\draw (158.12,137.82) node  {\includegraphics[width=44.27pt,height=43.96pt]{images/mix.png}};
\draw    (131.6,111.91) -- (190.63,111.91) -- (190.63,170.53) -- (131.6,170.53) -- cycle ;
\draw (161.12,141.22) node  {\includegraphics[width=44.27pt,height=43.96pt]{images/mix.png}};

\draw    (392,108.71) -- (451.03,108.71) -- (451.03,167.33) -- (392,167.33) -- cycle ;
\draw (421.52,138.02) node  {\includegraphics[width=44.27pt,height=43.96pt]{images/output.png}};

\draw  [fill={rgb, 255:red, 235; green, 199; blue, 204 }  ,fill opacity=1 ] (216.67,110.46) .. controls (216.67,109.07) and (217.79,107.95) .. (219.18,107.95) -- (226.7,107.95) .. controls (228.09,107.95) and (229.21,109.07) .. (229.21,110.46) -- (229.21,166.99) .. controls (229.21,168.38) and (228.09,169.5) .. (226.7,169.5) -- (219.18,169.5) .. controls (217.79,169.5) and (216.67,168.38) .. (216.67,166.99) -- cycle ;
\draw   (218,172.83) .. controls (217.99,177.5) and (220.32,179.84) .. (224.99,179.85) -- (284.65,179.97) .. controls (291.32,179.98) and (294.65,182.32) .. (294.64,186.99) .. controls (294.65,182.32) and (297.98,180) .. (304.65,180.01)(301.65,180.01) -- (364.32,180.14) .. controls (368.99,180.15) and (371.32,177.82) .. (371.33,173.15) ;

\draw     (230.21,138.95) -- (248.21,138.95) ;
\draw [shift={(250.21,138.95)}, rotate = 180] [color={rgb, 255:red, 0; green, 0; blue, 0 }  ][line width=0.75]    (10.93,-3.29) .. controls (6.95,-1.4) and (3.31,-0.3) .. (0,0) .. controls (3.31,0.3) and (6.95,1.4) .. (10.93,3.29)   ;
\draw     (265.21,138.95) -- (283.21,138.95) ;
\draw [shift={(285.21,138.95)}, rotate = 180] [color={rgb, 255:red, 0; green, 0; blue, 0 }  ][line width=0.75]    (10.93,-3.29) .. controls (6.95,-1.4) and (3.31,-0.3) .. (0,0) .. controls (3.31,0.3) and (6.95,1.4) .. (10.93,3.29)   ;
\draw     (301.21,138.95) -- (319.21,138.95) ;
\draw [shift={(321.21,138.95)}, rotate = 180] [color={rgb, 255:red, 0; green, 0; blue, 0 }  ][line width=0.75]    (10.93,-3.29) .. controls (6.95,-1.4) and (3.31,-0.3) .. (0,0) .. controls (3.31,0.3) and (6.95,1.4) .. (10.93,3.29)   ;
\draw     (332,140) -- (350,140) ;
\draw [shift={(352,140)}, rotate = 180] [color={rgb, 255:red, 0; green, 0; blue, 0 }  ][line width=0.75]    (10.93,-3.29) .. controls (6.95,-1.4) and (3.31,-0.3) .. (0,0) .. controls (3.31,0.3) and (6.95,1.4) .. (10.93,3.29)   ;

\draw     (372,140) -- (388,140) ;
\draw [shift={(390,140)}, rotate = 180] [color={rgb, 255:red, 0; green, 0; blue, 0 }  ][line width=0.75]    (10.93,-3.29) .. controls (6.95,-1.4) and (3.31,-0.3) .. (0,0) .. controls (3.31,0.3) and (6.95,1.4) .. (10.93,3.29)   ;

\draw     (189.57,66.43) -- (328.43,66.43) -- (328.65,103.5) ;
\draw [shift={(328.67,105.5)}, rotate = 269.65] [color={rgb, 255:red, 0; green, 0; blue, 0 }  ][line width=0.75]    (10.93,-3.29) .. controls (6.95,-1.4) and (3.31,-0.3) .. (0,0) .. controls (3.31,0.3) and (6.95,1.4) .. (10.93,3.29)   ;

\draw  [fill={rgb, 255:red, 217; green, 150; blue, 150 }  ,fill opacity=1 ] (251.67,110.79) .. controls (251.67,109.4) and (252.79,108.28) .. (254.18,108.28) -- (261.7,108.28) .. controls (263.09,108.28) and (264.21,109.4) .. (264.21,110.79) -- (264.21,167.32) .. controls (264.21,168.71) and (263.09,169.83) .. (261.7,169.83) -- (254.18,169.83) .. controls (252.79,169.83) and (251.67,168.71) .. (251.67,167.32) -- cycle ;
\draw  [fill={rgb, 255:red, 235; green, 199; blue, 204 }  ,fill opacity=1 ] (287,111.12) .. controls (287,109.74) and (288.12,108.61) .. (289.51,108.61) -- (297.04,108.61) .. controls (298.42,108.61) and (299.55,109.74) .. (299.55,111.12) -- (299.55,167.66) .. controls (299.55,169.04) and (298.42,170.17) .. (297.04,170.17) -- (289.51,170.17) .. controls (288.12,170.17) and (287,169.04) .. (287,167.66) -- cycle ;
\draw  [fill={rgb, 255:red, 178; green, 200; blue, 225 }  ,fill opacity=1 ] (322.67,110.12) .. controls (322.67,108.74) and (323.79,107.61) .. (325.18,107.61) -- (332.7,107.61) .. controls (334.09,107.61) and (335.21,108.74) .. (335.21,110.12) -- (335.21,166.66) .. controls (335.21,168.04) and (334.09,169.17) .. (332.7,169.17) -- (325.18,169.17) .. controls (323.79,169.17) and (322.67,168.04) .. (322.67,166.66) -- cycle ;
\draw  [fill={rgb, 255:red, 127; green, 179; blue, 239 }  ,fill opacity=1 ] (359.33,111.12) .. controls (359.33,109.74) and (360.46,108.61) .. (361.84,108.61) -- (369.37,108.61) .. controls (370.76,108.61) and (371.88,109.74) .. (371.88,111.12) -- (371.88,167.66) .. controls (371.88,169.04) and (370.76,170.17) .. (369.37,170.17) -- (361.84,170.17) .. controls (360.46,170.17) and (359.33,169.04) .. (359.33,167.66) -- cycle ;

\draw    (186.73,56.99) -- (121.53,57.08) -- (121.52,47.41) -- (186.72,47.32) -- cycle ;
\draw    (186.74,66.65) -- (121.55,66.75) -- (121.53,57.08) -- (186.73,56.99) -- cycle ;
\draw    (186.76,76.32) -- (121.56,76.41) -- (121.55,66.74) -- (186.74,66.65) -- cycle ;
\draw  [fill={rgb, 255:red, 245; green, 166; blue, 35 }  ,fill opacity=1 ] (186.77,85.99) -- (121.57,86.08) -- (121.56,76.41) -- (186.76,76.32) -- cycle ;
\draw    (186.78,95.65) -- (121.59,95.74) -- (121.57,86.08) -- (186.77,85.99) -- cycle ;
\draw    (186.72,47.32) -- (121.52,47.41) -- (121.5,37.74) -- (186.7,37.65) -- cycle ;

\draw (430.5,172) node [anchor=north west][inner sep=0.75pt]   [align=left] {{\tiny [T,K,2]}};
\draw (166.3,175.2) node [anchor=north west][inner sep=0.75pt]   [align=left] {{\tiny [T,K,2J]}};
\draw (359.33,161.62) node [anchor=north west][inner sep=0.75pt]  [font=\scriptsize,rotate=-270.2] [align=left] {{\scriptsize time-conv}};
\draw (125.67,28.2) node [anchor=north west][inner sep=0.75pt]   [align=left] {{\tiny {\fontfamily{pcr}\selectfont lookup table}}};
\draw (342.31,65.87) node [anchor=north west][inner sep=0.75pt]  [font=\scriptsize,rotate=-90.83]  {$E_{emb}$};
\draw (275.55,184.0) node [anchor=north west][inner sep=0.75pt]   [align=left] {{\footnotesize {\fontfamily{ptm}\selectfont L times}}};
\draw (287,161.24) node [anchor=north west][inner sep=0.75pt]  [font=\scriptsize,rotate=-270.2] [align=left] {{\scriptsize freq-conv}};
\draw (217,161.95) node [anchor=north west][inner sep=0.75pt]  [font=\scriptsize,rotate=-270.2] [align=left] {{\scriptsize freq-conv}};
\draw (252.34,160.62) node [anchor=north west][inner sep=0.75pt]  [font=\scriptsize,rotate=-270.2] [align=left] {{\scriptsize full-band}};
\draw (323.08,165.95) node [anchor=north west][inner sep=0.75pt]  [font=\scriptsize,rotate=-270.2] [align=left] {{\scriptsize self-attention}};

\end{tikzpicture}
        \caption{SpatialNet-based architecture}
        \label{fig:subfig_a}
    \end{subfigure}
    \hfill 

    \caption{Block diagram of the proposed \ac{RPC}-based target speaker extraction framework. \acfp{RPC} are used to guide the network toward the desired (target) speaker. The lower panel shows an implementation based on SpatialNet, while the upper panel shows one based on a U-Net architecture. In the SpatialNet implementation, the command embedding is concatenated with the feature representations, whereas in the U-Net implementation it is incorporated using \ac{FiLM} conditioning. These choices are not mandatory and are shown only as representative examples of how \ac{RPC} information can be integrated into existing architectures.}
    \label{fig:model}
\end{figure}

To guide the network toward the desired speaker, we associate each \ac{RPC} command with a learnable embedding vector using the same embedding mechanism described above. The command embedding is concatenated with the frame-level representations before the self-attention module in each narrow-band block. Consequently, the self-attention mechanism jointly processes the acoustic representations and the command embedding, enabling the network to model dependencies between the \ac{RPC} instruction and all time-frequency regions of the input signal. This allows the command information to guide the attention mechanism toward the desired speaker, as illustrated in Fig.~\ref{fig:subfig_a}.

\subsection{Features and Objective Function}
In this study, we employed the real and imaginary (RI) components of the \ac{STFT} as both input features and output targets for the model. These components were concatenated along the channel dimension. To reconstruct the signal in the time domain, the inverse STFT was applied to the model's output.
We optimize the proposed extraction model using the time-domain scale-invariant \ac{SI-SDR} loss \cite{le2019sdr}, a widely adopted objective for \ac{BSS}. For each training mixture, both speakers are considered in turn as the target speaker, conditioned on their corresponding {\ac{RPC}} sample.
\rev{The \ac{SI-SDR} between a reference signal $s$ and its estimate $\hat{s}$ is defined as~\cite{le2019sdr}}
\begin{equation}
\text{SI-SDR}\left( s,\hat{s} \right)= 10 \log_{10} \left( \frac{\norm{\frac{\langle {\hat{s},s} \rangle}{\langle {s,s} \rangle} s}^2}{\norm{\frac{\langle {\hat{s},s} \rangle}{\langle {s,s} \rangle} s-\hat{s}}^2} \right).
\end{equation}
The resulting loss is given by:
\begin{equation}
\mathcal{L} = \rev{-}\sum_{i=1}^{2} \text{SI-SDR}\left(\mathbf{s}_i,\hat{\mathbf{s}}_i\right)\rev{,}
\end{equation}
\rev{where $\mathbf{s}_i$ denotes the reverberant image of speaker $i$ at the reference microphone and $\hat{\mathbf{s}}_i$ is the network output when conditioned on the command of speaker $i$. The loss is averaged over the three spatial variants of each mixture (Sec.~\ref{sec:datasets}). In the three-speaker experiments, a single speaker per mixture serves as the target, and the sum in the loss reduces to a single term. Separate models are trained for each number of speakers ($Q=2$ and $Q=3$, with eight and nine commands, respectively) and for each array size (two and four microphones).}

\section{Experimental Study}
\label{sec:exp}
This section presents the experimental evaluation of the proposed method. We begin by describing the datasets and training configuration, followed by the competing methods used as baselines. We then report results under different spatial configurations, microphone array sizes, and speaker counts, using both simulated and real room impulse responses.

\subsection{Datasets}
\rev{\label{sec:datasets}}
To generate training, validation, and test data, we synthetically created 20,000 training examples, 1,000 validation examples, and 1,000 test examples. The clean speech signals were sampled from the Librispeech dataset, and the noise samples were sampled from the DNS dataset. \rev{Training, validation, and test speakers are drawn from the disjoint LibriSpeech training, development, and test subsets, respectively, and the DNS noise files are split 80/10/10\,\% among the three sets.} The \ac{SNR} was randomly chosen in the range of 5 to 20 dB. To simulate reverberant environments, we used a simulation framework based on the Image method \cite{habets2006room}, in which the microphone array's location and azimuth (relative to the wall) were randomly assigned for each room. The reverberation time was randomly selected from the range 0.2 to 0.8 seconds. A \ac{ULA} comprising four microphones, with an 8 cm spacing between adjacent microphones, was used. The minimum distance from the wall was set to 0.5 m.
\rev{The room dimensions are drawn uniformly in $[6,10]\times[6,10]\times3$~m for each example, and the array is placed at a random position, such that all sources are at least 0.5~m from the walls. All sources are placed at a height of 1.5~m, in the half-plane in front of the array, on a $2^\circ$ \ac{DOA} grid; the distances $R$ and \acp{DOA} $\theta$ are measured in the horizontal plane with respect to the array center. The two-microphone configuration uses the two middle microphones of the same array (8~cm spacing).}

\rev{Each example comprises a \emph{primary} speaker, a \emph{second}, different speaker, and a directional background-noise source. The primary speaker is placed at a distance $R_1\sim\mathcal{U}[1.5,2]$~m from the array center and at a random \ac{DOA}. The noise source is placed at a random distance and at a \ac{DOA}. The same utterance of the second speaker is then rendered at three different positions, one per spatial configuration targeted by our model:
(i) both a large radial gap, $\psi>0.5$~m, and a large \ac{DOA} gap, $\phi>10^\circ$;
(ii) a large radial gap, $\psi>0.5$~m, and a small \ac{DOA} gap, $\phi\in[2^\circ,10^\circ]$, i.e., both speakers are located in practically the same direction; and
(iii) a small radial gap, $\psi<0.5$~m, and a large \ac{DOA} gap, $\phi>10^\circ$.
 These three configurations define the three test sets (i)--(iii) used throughout this section. Since the absolute positions are random, the absolute location of a speaker does not determine its command; for example, the second speaker can be located between 0.5 and 3~m from the array, and can be either the closer or the farther speaker.}
During training, these spatial variants are used to generate three distinct mixtures and three corresponding {command} utterances per speaker, ensuring the model is exposed to all spatial dimensions of the task.

\rev{For the three-speaker dataset, a third speaker is added to each of the three spatial configurations in the same manner as the second speaker, and the primary speaker serves as the target. The commands follow Table~\ref{table:speaker_positions_3}.}

\subsection{Algorithm Settings}

To prepare the data, the speech and noise signals were first resampled to 8 kHz. The signals were then converted to the frequency domain by computing the \ac{STFT} with a 256-sample frame size, a $50\%$ overlap between frames, and a Hamming window. Only the first 129 frequency bins were processed due to the symmetry of the DFT. The neural network was trained on 4-second signal segments using the AdamW optimizer~\cite{loshchilov2017decoupled}, with a learning rate of 0.0001 and a batch size of $B = 10$. Since each training batch includes all three spatial cases and each speaker is alternately used, the effective batch size \rev{in the two-speaker case} is $B_{\text{full}} = B * 6 = 60.$ This setup ensures comprehensive exposure to spatial diversity and balanced participation of each speaker across both separation and extraction objectives. \rev{In the three-speaker case, only the primary speaker serves as the target in each mixture, and the effective batch size is therefore $B_{\text{full}} = 3B = 30$.}


\subsection{Competing Methods}
 For comparison with existing methods, we considered FasNet \cite{luo2019fasnet} and SpatialNet \cite{quan2024spatialnet}, which are representative multi-channel speech separation frameworks. It is important to note that these methods address slightly different tasks: in separation, the model aims to recover all speakers from the mixture, whereas in extraction, it must also identify and recover only the target speaker. Nevertheless, since SpatialNet is a strong representative of spatial speech separation methods, we adapted it to the extraction setting by conditioning it on the target speaker embedding and retrained it on our dataset to serve as a competitive baseline. \rev{Since the separation models (FasNet and SpatialNet) do not select the target, their outputs are assigned to the speakers using the permutation that maximizes the \ac{SI-SDR}, as in \ac{PIT}. This oracle selection favors the separation baselines over the extraction methods, which must select the target on their own.} 

We also examine the \ac{SSF} framework \cite{tesch2023spatially}, a microphone-array speaker-extraction method that uses the target speaker’s \ac{DOA} as a reference. We adopt its architecture and replace the original objective with the \ac{SI-SDR} loss, thereby aligning it with our proposed framework.

To assess the effectiveness of the proposed relative-reference approach, we compared our U-Net model with a variant that uses the target speaker's absolute \ac{DOA} for enrollment instead of a ratio-based command. To this end, we introduced a larger lookup table that learns a dedicated embedding for each \ac{DOA} \rev{on the $2^\circ$ grid (90 entries)}. Aside from this change, the architecture remained identical to the original model. \rev{As for all other models, separate \ac{DOA}-based models are trained for the two- and four-microphone arrays.}

Finally, we compared our model with an oracle \ac{MVDR} beamformer \cite{souden2010optimal}. The \ac{MVDR} beamformer is designed to estimate the target signal with minimal distortion while suppressing noise and interfering speakers. Letting $\mathbf{g}_d(k)$ denote the \ac{MVDR} beamformer at frequency bin $k$, the estimated signal can be expressed as:
\begin{equation}
\hat{s}_d(k,t) = \mathbf{g}^{\text{H}}_d(k) \cdot \mathbf{x}(t,k),
\end{equation}
where $(\cdot)^\textrm{H}$ denotes the Hermitian transpose.
The \ac{MVDR} weights are defined as
\begin{equation}
\label{eq:mvdr}
\mathbf{g}_d(k) = \frac{\Phi^{-1}_{nn}\Phi_{ss}}{\mathrm{Tr}(\Phi^{-1}_{nn}\Phi_{ss})} \mathbf{u},
\end{equation}
where $\Phi_{nn}$ and $\Phi_{ss}$ represent the spatial covariance matrices of the noise (including the interfering speaker) and the desired speaker, respectively. The vector $\mathbf{u}$ is a one-hot vector specifying the reference channel. For the oracle setting considered here, we assume access to the noise and interfering sources in order to compute the \ac{MVDR} weights. 

\subsection{Results}
\rev{\label{sec:results}}
The \ac{SISDR} performance of the proposed models is reported in the following tables\rev{, together with the \ac{STOI}~\cite{taal2011algorithm} of the same outputs, computed at 8~kHz with respect to the same reference signal}. To comprehensively evaluate our approach, we tested it under three different scenarios: (i) large gaps in both \ac{DOA} and radius, (ii) small \ac{DOA} gap with large radius gap, and (iii) small radius gap with large \ac{DOA} gap\rev{, as defined in Sec.~\ref{sec:datasets}. The same indices (i)--(iii) are used in all tables. The two-speaker results for four and two microphones are summarized in Table~\ref{table:results}. Entries marked by ``x'' are not reported: in test set (ii), the two speakers are located in practically the same direction ($\phi\le10^\circ$), so a \ac{DOA}-based clue (SSF and U-Net DOA) cannot unambiguously identify the target.}

\rev{We first discuss the four-microphone results in Table~\ref{table:results}.} The oracle \ac{MVDR} achieved similar \ac{SISDR} scores in the large-\ac{DOA}-gap scenarios, while exhibiting a modest degradation in the small-\ac{DOA}-gap case. Among the separation methods, FasNet performed significantly worse than the other evaluated approaches. By contrast, SpatialNet consistently achieved strong results across all test conditions.

Among the extraction methods, the \ac{RPC}-based SpatialNet model outperformed even its separation counterpart. This is particularly remarkable because speaker extraction is inherently more challenging than speaker separation, requiring both identification and isolation of a specific target speaker. Despite this added complexity, the extraction model achieved superior performance, demonstrating the effectiveness of the proposed \ac{RPC} framework.

Finally, the \ac{RPC}-based U-Net variants consistently outperformed their absolute \ac{DOA}-based counterparts across all test sets\rev{ (by 0.7--1.1~dB with four microphones and 0.8--1.2~dB with two microphones, Table~\ref{table:results})}, demonstrating the benefit of using \ac{RPC} commands in conjunction with the corresponding \ac{DOA} information. Furthermore, \ac{DOA}-based enrollment alone is insufficient to distinguish between speakers in the same direction, as it provides no information about their relative distances from the microphone array.

\begin{figure}[t]
\centering
\includegraphics[width=\columnwidth]{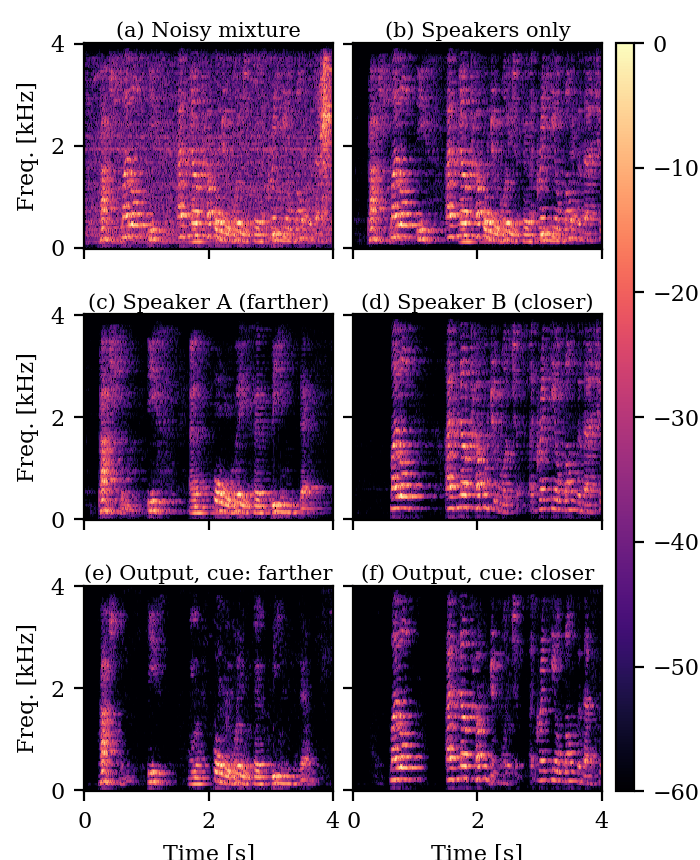}
\caption{\rev{Spectrograms (reference microphone) of a test mixture from test set (ii), processed by the four-microphone U-Net RPC model. Speaker A is at 1.68~m and $62^\circ$, speaker B at 0.70~m and $56^\circ$; $T_{60}=0.79$~s, \ac{SNR} $=6$~dB, \ac{SIR} $=-0.6$~dB. (a) Noisy mixture. (b) Speech-only mixture (ideal noise suppression). (c), (d) Reverberant references of speakers A and B. (e) Output with the command ``farther'': \ac{SI-SDR} of 10.1~dB w.r.t.\ A and $-26.5$~dB w.r.t.\ B. (f) Output with the command ``closer'': 10.2~dB w.r.t.\ B and $-21.8$~dB w.r.t.\ A.}}
\label{fig:spec}
\end{figure}

\rev{Since the mixtures also contain background noise, part of the improvement could, in principle, stem from noise suppression rather than from target extraction. This is not the case: an ideal noise suppressor that removes the noise but keeps both speakers yields an \ac{SI-SDR} of only 0.0~dB, i.e., only 0.75~dB above the unprocessed mixture, whereas the \ac{RPC}-based models reach 9.0--14.2~dB (Table~\ref{table:results}). Fig.~\ref{fig:spec} illustrates this for a mixture from test set (ii), in which the two speakers are located in practically the same direction. Processing the same mixture with the commands ``farther'' and ``closer'' yields two different outputs, each containing only the commanded speaker (panels (e) and (f)), whereas noise suppression alone would retain both speakers (panel (b)).}

We further evaluated our models in scenarios with a limited number of microphones to assess the impact on performance. In general, increasing the number of microphones provides richer spatial information that can be exploited to improve separation quality. Conversely, when only two microphones are available, models that rely heavily on spatial cues face a more challenging task.

As shown in Table~\ref{table:results}, the performance of the \ac{MVDR} beamformer deteriorates when restricted to two microphones, owing to the reduced degrees of freedom in the minimization. Among the separation methods, SpatialNet continues to exhibit strong performance despite the limited spatial information. FasNet also remains relatively robust, although its performance is slightly lower than that achieved with the four-microphone configuration.

For the extraction-based methods, our model exhibits \rev{only a small performance degradation of 0.2--0.4~dB} relative to the four-microphone configuration\rev{, similar to that of the U-Net DOA model}. Nevertheless, it consistently achieves the best results across all evaluated scenarios. It is also noteworthy that, unlike the four-microphone setting, the extraction variant of SpatialNet performs worse than its separation counterpart, despite sharing the same underlying architecture.

This contrasting behavior can be attributed to the fundamental differences between speaker extraction and speaker separation. Whereas separation models aim to recover all speakers present in the mixture and can leverage both spatial and spectral cues, the \ac{RPC}-based extraction model is trained to primarily rely on spatial information to identify and isolate the target speaker. Consequently, its performance is closely tied to the quality of the available spatial cues. When the spatial information is rich and reliable, the model can effectively focus on and extract the desired speaker. However, with only a small number of microphones, the spatial information becomes less informative, leading to a more pronounced performance degradation. Despite this limitation, the proposed method remains superior to all competing extraction methods. \rev{The per-mixture analysis in Sec.~\ref{sec:sensitivity} shows that the degradation of SpatialNet RPC is not caused by selecting the wrong speaker (the correct speaker is extracted in more than 98\,\% of the cases), but by a lower extraction quality across all conditions, with the largest drop for small \ac{DOA} gaps (Fig.~\ref{fig:conditions}). Since the U-Net-based RPC model degrades by only 0.2--0.4~dB, this behavior appears to be specific to the combination of the architecture and the conditioning mechanism rather than inherent to the \ac{RPC}. A detailed study of the interplay between the conditioning mechanism, the architecture, and the array size is beyond the scope of this paper.}








\begin{table}[ht]
\caption{\ac{SI-SDR}\rev{ and \ac{STOI}} results for 2- and 4-microphone \ac{ULA}\rev{. Test sets: (i) large \ac{DOA} and radial gaps; (ii) small \ac{DOA} gap, large radial gap; (iii) small radial gap, large \ac{DOA} gap. x: not applicable. Each entry: \ac{SI-SDR} [dB] / \ac{STOI}.}}
\label{table:results}
\begin{center}
\resizebox{\columnwidth}{!}{%
\begin{tabular}{@{}lcccccc@{}}
\toprule
& \multicolumn{3}{c}{\textbf{4 Microphones}} & \multicolumn{3}{c}{\textbf{2 Microphones}} \\
\cmidrule(lr){2-4} \cmidrule(lr){5-7}
\textbf{Model / Test Set} & (i) & (ii) & (iii) & (i) & (ii) & (iii) \\ \midrule
Unprocessed & -0.75/\rev{0.598} & -0.76/\rev{0.598} & -0.75/\rev{0.598} & -0.75/\rev{0.599} & -0.76/\rev{0.599} & -0.75/\rev{0.598} \\ \midrule
Oracle MVDR & 8.80/\rev{0.839} & 8.07/\rev{0.822} & 8.80/\rev{0.840} & 5.44/\rev{0.737} & 4.57/\rev{0.714} & 5.47/\rev{0.737} \\ \midrule
FasNet \cite{luo2019fasnet} & 7.47/\rev{0.733} & 6.90/\rev{0.678} & 7.60/\rev{0.735} & 6.80/\rev{0.744} & 5.40/\rev{0.711} & 6.90/\rev{0.745} \\
SpatialNet \cite{quan2024spatialnet} & 13.20/\rev{0.922} & 12.20/\rev{0.912} & 13.20/\rev{0.922} & 12.39/\rev{0.875} & 11.80/\rev{0.866} & 12.43/\rev{0.877} \\ \midrule
SpatialNet RPC & 14.20/\rev{0.911} & 13.40/\rev{0.901} & 14.10/\rev{0.909} & 10.71/\rev{0.848} & 9.00/\rev{0.811} & 10.61/\rev{0.846} \\
SSF \cite{tesch2023spatially} & 9.10/\rev{0.812} & x & 9.05/\rev{0.812} & 9.09/\rev{0.818} & x & 9.01/\rev{0.817} \\
Unet DOA & 9.90/\rev{0.847} & x & 10.29/\rev{0.849} & 9.69/\rev{0.827} & x & 9.85/\rev{0.832} \\
Unet RPC & 11.02/\rev{0.863} & 9.75/\rev{0.835} & 11.00/\rev{0.863} & 10.86/\rev{0.858} & 9.39/\rev{0.826} & 10.67/\rev{0.854} \\
\bottomrule
\end{tabular}}
\end{center}
\end{table}
The results for the three-speaker configuration, presented in Table~\ref{table:results_3_speakers}, demonstrate that the proposed model maintains stable performance as the number of speakers increases. Notably, the model remains effective even when the number of speakers exceeds the number of microphones, i.e., when the available degrees of freedom are lower than the number of speakers in the mixture. \rev{When the number of microphones is reduced from four to two, the performance of the U-Net RPC model decreases by only about 0.5~dB, and that of SpatialNet RPC by only 0.3--0.4~dB,} highlighting the robustness of the proposed approach. 

In contrast, reducing the number of microphones from four to two results in larger performance degradation for the \ac{DOA}-based model\rev{ (1.4--1.5~dB)} and an even more pronounced decline for the oracle \ac{MVDR} beamformer\rev{ (about 4~dB)}. These findings further demonstrate the robustness of the proposed relative-reference approach, particularly in challenging underdetermined scenarios where spatial information is limited and conventional spatial filtering methods become less effective.






\begin{table}[ht]
\caption{SI-SDR\rev{ and \ac{STOI}} results for 3 speakers with 4 and 2 microphones\rev{. Test sets (i)--(iii) as in Table~\ref{table:results}; x: not applicable. Each entry: \ac{SI-SDR} [dB] / \ac{STOI}.}}
\label{table:results_3_speakers}
\begin{center}
\resizebox{\columnwidth}{!}{%
\begin{tabular}{@{}lcccccc@{}}
\toprule
& \multicolumn{3}{c}{\textbf{4 Microphones}} & \multicolumn{3}{c}{\textbf{2 Microphones}} \\
\cmidrule(lr){2-4} \cmidrule(lr){5-7}
\textbf{Model / Test Set} & (i) & (ii) & (iii) & (i) & (ii) & (iii) \\ \midrule
Unprocessed & -3.84/\rev{0.469} & -3.87/\rev{0.469} & -3.84/\rev{0.468} & -3.84/\rev{0.469} & -3.87/\rev{0.469} & -3.84/\rev{0.468} \\ \midrule
Oracle MVDR & 5.70/\rev{0.747} & 4.99/\rev{0.725} & 5.65/\rev{0.746} & 1.55/\rev{0.614} & 1.03/\rev{0.595} & 1.48/\rev{0.612} \\
SpatialNet RPC & 6.70/\rev{0.768} & 5.70/\rev{0.736} & 6.07/\rev{0.751} & 6.40/\rev{0.772} & 5.40/\rev{0.744} & 5.70/\rev{0.754} \\
Unet DOA & 5.34/\rev{0.740} & x & 5.10/\rev{0.733} & 3.98/\rev{0.689} & x & 3.56/\rev{0.680} \\
Unet RPC & 6.31/\rev{0.768} & 4.77/\rev{0.719} & 5.90/\rev{0.759} & 5.87/\rev{0.751} & 4.20/\rev{0.699} & 5.42/\rev{0.740} \\
\bottomrule
\end{tabular}}
\end{center}
\end{table}

We also conducted experiments using real recorded \acp{RIR} obtained from \cite{hadad2014multichannel}. The recordings were made at two possible source–microphone distances: 1 meter and 2 meters. The \acp{DOA} were sampled at 15° intervals between 0° and 180°. {Here, scenarios with small \ac{DOA} gaps mean the \acp{DOA} are identical, differing only in their radial distance.} In \cite{hadad2014multichannel}, three reverberation times were provided: 160 ms, 360 ms, and 610 ms. Our evaluation focused on the 360 ms case. The results, presented in Table~\ref{table:results_real_data}, indicate that the proposed method achieves performance comparable to that obtained with simulated data across all tested architectures and microphone numbers.

\begin{table}[ht]
\caption{SI-SDR \rev{and STOI} results for real data with 4 and 2 microphones.}
\label{table:results_real_data}
\begin{center}
\resizebox{\columnwidth}{!}{%
\begin{tabular}{@{}lcccccc@{}}
\toprule
& \multicolumn{3}{c}{\textbf{4 Microphones}} & \multicolumn{3}{c}{\textbf{2 Microphones}} \\
\cmidrule(lr){2-4} \cmidrule(lr){5-7}
\textbf{Model / Test Set} & (i) & (ii) & (iii) & (i) & (ii) & (iii) \\ \midrule
Unprocessed    & -0.75/\rev{0.598}  & -0.76/\rev{0.598}  & -0.62/\rev{0.541}  & -0.75/\rev{0.598}  & -0.76/\rev{0.598}  & -0.62/\rev{0.541}  \\ \midrule
SpatialNet RPC & 12.35/\rev{0.914} & 7.65/\rev{0.741}   & 12.97/\rev{0.901}  & 9.20/\rev{0.823}   & 5.50/\rev{0.713}   & 10.50/\rev{0.885}  \\
Unet RPC       & 9.39/\rev{0.829}   & 5.24/\rev{0.736}   & 9.70/\rev{0.853}   & 8.80/\rev{0.804}   & 4.70/\rev{0.721}   & 8.44/\rev{0.766}   \\
\bottomrule
\end{tabular}}
\end{center}
\end{table}

\subsection{\rev{Robustness to Diffuse Babble Noise}}
\label{sec:babble}
\begin{figure}[t]
\centering
\includegraphics[width=0.85\columnwidth]{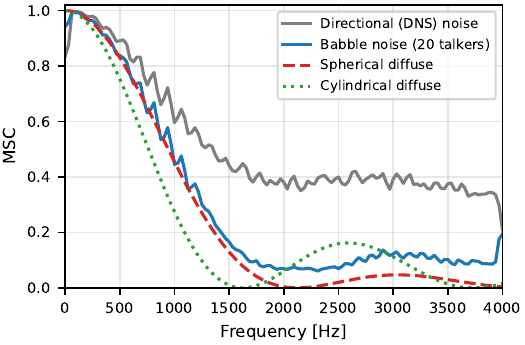}
\caption{\rev{Magnitude-squared coherence between two adjacent microphones ($d=8$~cm), averaged over the test set, for the directional (DNS) noise and the 20-talker babble noise, compared with the theoretical spherically and cylindrically isotropic (diffuse) noise fields.}}
\label{fig:babble_coherence}
\end{figure}

{\color{revgreen}
In all the experiments above, the background noise is a single directional source, i.e., $\mathbf{v}(t,k)=\hn \cdot n(t,k)$ in \eqref{eq:model}. To examine how the \ac{RPC}-based models behave with diffuse noise, we constructed a babble-noise version of the static two-speaker test set. For each test mixture, the directional noise was replaced by babble noise, while the target and interfering speech signals, the room, and the array were kept unchanged. The babble noise was generated by placing 20 different talkers on a circle centered at the microphone array, at the height of the speakers. The talkers were uniformly spaced over $360^\circ$ (with a random rotation of the circle). The \acp{RIR} of the 20 talkers were simulated using the image method~\cite{habets2006room} in the same room, with the same reverberation time and microphone-array position as the corresponding test mixture, and the 20 reverberant talker signals were summed to form $\mathbf{v}(t,k)$. The \ac{SNR} between the two speakers and the babble noise was drawn uniformly from 10--20~dB. Neither model was retrained or fine-tuned, i.e., the models were never exposed to diffuse noise during training.

Fig.~\ref{fig:babble_coherence} depicts the magnitude-squared coherence (MSC) between two adjacent microphones ($d=8$~cm), averaged over the test set. The coherence of the babble noise closely follows the theoretical coherence of a spherically isotropic noise field, $\mathrm{sinc}^2(2\pi f d/c)$, whereas the directional noise remains highly coherent across the entire frequency band. The babble noise is therefore a good approximation of a diffuse noise field, and, unlike the directional noise, it cannot be cancelled by spatial nulling.

The results are summarized in Table~\ref{table:results_babble}. All \ac{RPC}-based models maintain their performance in the presence of diffuse babble noise. The U-Net RPC model degrades by at most 0.25~dB with both four and two microphones, and the two-microphone SpatialNet RPC degrades by at most 0.3~dB. The largest degradation, of approximately 0.9~dB, is observed for the four-microphone SpatialNet RPC, which nevertheless remains the best-performing model. The oracle \ac{MVDR} beamformer exhibits similar trends, with a minor degradation for four microphones and a minor improvement for two microphones, where the directional noise source constitutes an additional point source in an already under-determined setting.

Importantly, although the babble noise consists of speech, the models did not confuse the background talkers with the speakers to which the command refers, and the relative ordering of the three test scenarios is preserved. The \ac{RPC} specifies the target with respect to the dominant, spatially coherent interfering speaker, whereas each babble talker is weak and the babble mixture lacks a distinct spatial signature. These results indicate that the proposed relational conditioning remains reliable when the directional-noise assumption is violated.
}

\begin{table}[ht]
\color{revgreen}
\caption{\ac{SI-SDR} and \ac{STOI} results for directional (Dir.) and diffuse babble (Bab.) noise with 4- and 2-microphone \ac{ULA}. The models were trained with directional noise only. Each entry: \ac{SI-SDR} [dB] / \ac{STOI}.}
\label{table:results_babble}
\begin{center}
\resizebox{\columnwidth}{!}{%
\begin{tabular}{@{}llcccccc@{}}
\toprule
& & \multicolumn{3}{c}{\textbf{4 Microphones}} & \multicolumn{3}{c}{\textbf{2 Microphones}} \\
\cmidrule(lr){3-5} \cmidrule(lr){6-8}
\textbf{Model} & \textbf{Noise} & (i) & (ii) & (iii) & (i) & (ii) & (iii) \\ \midrule
\multirow{2}{*}{Unprocessed} & Dir. & -0.75/0.598 & -0.76/0.598 & -0.75/0.598 & -0.75/0.599 & -0.76/0.599 & -0.75/0.598 \\
 & Bab. & -0.39/0.610 & -0.40/0.610 & -0.39/0.610 & -0.39/0.610 & -0.39/0.610 & -0.39/0.610 \\ \midrule
\multirow{2}{*}{Oracle MVDR} & Dir. & 8.81/0.839 & 8.07/0.822 & 8.81/0.840 & 5.44/0.737 & 4.57/0.714 & 5.47/0.737 \\
 & Bab. & 8.67/0.824 & 7.95/0.807 & 8.67/0.825 & 5.63/0.737 & 4.78/0.715 & 5.66/0.738 \\ \midrule
\multirow{2}{*}{SpatialNet RPC} & Dir. & 14.20/0.911 & 13.40/0.901 & 14.10/0.909 & 10.71/0.848 & 9.00/0.811 & 10.61/0.846 \\
 & Bab. & 13.29/0.891 & 12.51/0.880 & 13.28/0.890 & 10.42/0.835 & 8.81/0.801 & 10.42/0.835 \\ \midrule
\multirow{2}{*}{Unet RPC} & Dir. & 11.02/0.863 & 9.75/0.835 & 11.00/0.863 & 10.86/0.858 & 9.39/0.826 & 10.67/0.854 \\
 & Bab. & 10.88/0.855 & 9.69/0.829 & 10.76/0.852 & 10.67/0.848 & 9.48/0.822 & 10.50/0.843 \\
\bottomrule
\end{tabular}}
\end{center}
\end{table}

\subsection{\rev{Relation versus Region}}
\rev{\label{sec:relation_vs_region}}
\rev{\textit{The concern.} One could argue that the network does not learn the relation between the speakers, but only a fixed partition of the room. For example, it could interpret ``right'' as ``on the right side of the array broadside'' ($\theta<90^\circ$), and ``closer'' as ``closer than some fixed distance''. Such a fixed-region interpretation fails whenever both speakers are in the same region. If the target is at $100^\circ$ and the interferer at $130^\circ$, both are on the left side of the array, yet the target is \emph{to the right of the interferer}. Similarly, if the target is at 1.6~m and the interferer at 2.3~m, both are farther than 1.5~m, yet the target is the \emph{closer} speaker.}


\rev{\textit{What the models do.} We count an output as correct if it is closer, in \ac{SI-SDR}, to the commanded speaker than to the other speaker. Over all two-speaker test mixtures, the four \ac{RPC}-based models (two architectures, two array sizes) select the correct speaker in 98.4--100\,\% of the cases). Crucially, this rate remains just as high in the mixtures in which a fixed region would fail. When both speakers are on the same side of the broadside, the correct speaker is selected in 98.9--99.8\,\% of the cases, compared with 100\,\% when they are on opposite sides. For distance, the rate is 98.9--99.9\,\% both when the speakers are on the same side of a 1.75-m threshold and when they are on opposite sides, and the \ac{SI-SDR} differs by at most 0.2~dB between the two groups. The only noticeable difference is a 0.6--1.4~dB lower \ac{SI-SDR} when both speakers are on the same side of the broadside; it is explained by the smaller \ac{DOA} gaps in these mixtures (Fig.~\ref{fig:conditions}), not by selecting the wrong speaker. We conclude that the models resolve the relation between the speakers rather than a fixed partition of the room. A direct comparison with a model trained on region labels is left for future work.}

\subsection{\rev{Sensitivity to Spatial Gaps, Reverberation, and SIR}}
\begin{figure*}[t]
\centering
\includegraphics[width=\textwidth]{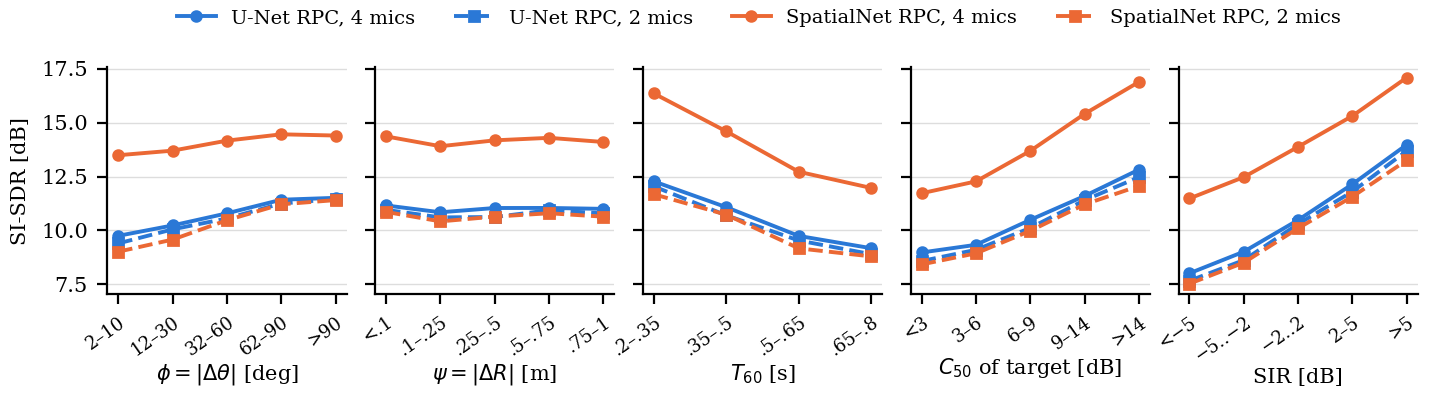}
\caption{\rev{\ac{SI-SDR} of the \ac{RPC}-based models (two-speaker test sets) as a function of the \ac{DOA} gap $\phi$ (test sets (i) and (ii)), the radial gap $\psi$ (test sets (i) and (iii)), the reverberation time $T_{60}$, the clarity index $C_{50}$ of the target, and the \ac{SIR}. Each point is the mean over all outputs within the bin.}}
\label{fig:conditions}
\end{figure*}

\rev{\label{sec:sensitivity}}
\rev{Fig.~\ref{fig:conditions} presents the per-mixture \ac{SI-SDR} of the \ac{RPC}-based models in the two-speaker test sets as a function of the \ac{DOA} gap $\phi$ (test sets (i) and (ii), where $\psi\ge0.5$~m), the radial gap $\psi$ (test sets (i) and (iii), where $\phi>10^\circ$), the reverberation time, the clarity index $C_{50}$ of the target at the reference microphone, and the \ac{SIR}. Note that the \ac{DOA} grid resolution is $2^\circ$, so all \ac{DOA} gaps are integer multiples of the grid resolution, and the smallest gap is $2^\circ$. The performance increases monotonically with $\phi$, from 9.0--13.5~dB for $\phi\le10^\circ$ to 11.4--14.4~dB for $\phi>90^\circ$; even when the speakers are in practically the same direction, the radial relation suffices for extraction. In contrast, the performance is almost independent of $\psi$ (variations within 0.5~dB), including $\psi<0.1$~m, where the speakers are practically equidistant and the angular relation is used. As expected, the performance degrades with increasing reverberation, by 2.9--4.4~dB between $T_{60}\in[0.2,0.35]$~s and $T_{60}\in[0.65,0.8]$~s, and increases with $C_{50}$. Finally, the \ac{SI-SDR} increases by about 6~dB from $\text{SIR}<-5$~dB to $\text{SIR}>5$~dB; however, the correct speaker is extracted in 97.5--100\,\% of the cases even when the target is more than 5~dB weaker than the interferer, indicating that the relative power of the speakers does not confuse the target selection within the tested range.}

\section{Moving Speakers}
\rev{\label{sec:moving}}
To assess the model's robustness in more complex environments, we simulated dynamic scenarios where speaker positions change over time. These moving-source conditions represent a significant increase in task difficulty compared to static setups. Our goal is to determine whether the model can isolate the target speaker when both the radial separation (“gap distance”) and the angular separation (i.e., \ac{DOA} difference) vary over time, despite the model being trained exclusively in static scenarios.

\paragraph{Dynamic scenarios.}
Assuming a scenario with two concurrent speakers, we define the time-varying distance and \ac{DOA} differences as
\begin{equation}
\psi(t) \triangleq \left| d_1(t)-d_2(t) \right|,\qquad
\phi(t) \triangleq \left| \theta_1(t)-\theta_2(t) \right|,
\end{equation}
where $d_i(t)$ and $\theta_i(t)$ denote the instantaneous radial distance and \ac{DOA} of speaker $i$, respectively. Let $\psi_{\min}$ and $\phi_{\min}$ be the thresholds defined above.

To test this setting, we constructed dynamic test sets divided into three categories:
\begin{enumerate}
\item \textit{{DOA}-only information}: $\psi(t) < \psi_{\min}$ and $\phi(t) \ge \phi_{\min}$ for all $t$.

\item \textit{{Distance}-only information}: $\psi(t) \ge \psi_{\min}$ and $\phi(t) < \phi_{\min}$ for all $t$.

\item \textit{combined DOA and {Distance} information}: $\psi(t) \ge \psi_{\min}$ and $\phi(t) \ge \phi_{\min}$ for all $t$.

\end{enumerate}
Note that while the individual \acp{DOA} and gap distances may vary over time, the relative distance between the two speakers remains valid. These configurations enable a systematic evaluation of the model's robustness to time-varying speaker positions and orientations.


\paragraph{Data Generation.}
To evaluate the performance of \ac{RPC}-based extraction schemes in dynamic scenarios, we simulated speaker movements along linear, arc-shaped, and circular trajectories. These dynamic scenarios were excluded from the training data. Each speaker independently followed one of the three trajectory types, subject to the constraint that the \ac{RPC} command remained constant throughout the utterance. In addition, each scenario included a stationary additive noise source that remained fixed throughout the simulation.
\begin{itemize}
    \item \textbf{Linear movement:} For each trajectory, a starting position within the room, a direction of motion, and a velocity (in meters per second) were sampled at random, with the only constraint being that the trajectory did not intersect the room walls.
    
   \item \textbf{Arc movement:} An initial direction of arrival (DOA), a terminal DOA, and a radius were randomly selected. The velocity was then computed from the signal duration and the total arc length traveled.
    
   \item \textbf{Circular movement:} A circular trajectory was randomly generated by selecting its radius and an initial DOA. The speaker then moved along the trajectory at a constant angular velocity. The angular velocity was computed from the signal duration and the total circumference traveled.
\end{itemize}
For all movement types, additional constraints were imposed to prevent trajectories from overlapping while preserving each category's definition.

\paragraph{Results.} The tracking results are summarized in Table~\ref{table:results_moving}. We first evaluated the performance of the SpatialNet separation model under dynamic conditions. As expected, its performance degrades relative to the static scenarios. When a sufficient \ac{DOA} separation exists between the speakers, SpatialNet maintains reasonable performance, achieving \ac{SISDR} scores of 6--9~dB. However, as the angular separation decreases, its performance deteriorates substantially.

In contrast, the proposed method, which extracts the target speaker based on relative spatial cues, remains robust to speaker motion as long as the relative positions of the target and interfering speakers are preserved. This result highlights the advantage of exploiting relative spatial relationships rather than relying solely on absolute speaker locations.

Our model conditions the mixture embedding on the \ac{RPC} and applies a global self-attention mechanism across the entire signal. Under speaker motion, however, long-range dependencies can be unnecessary, or even detrimental, because the relational label is defined instantaneously between the speakers at each time frame, rather than across their full trajectories. For example, if the task is to extract the speaker to the right, this relation is meaningful only with respect to the interferer’s position at the same time instant. A target position at one time may not be to the right of the interferer’s position at another, non-synchronous time instant. Therefore, aggregating evidence across distant frames may introduce spurious associations that are inconsistent with the instantaneous \ac{RPC}. An illustration is provided in Fig.~\ref{fig:speaker-trajectory}. One speaker follows a straight path while the other moves along an arc; although the target remains to the right of the interferer at every time instant, this relation does not necessarily hold between points sampled from different times along their trajectories. Consequently, global attention may exploit misleading cross-temporal dependencies.

To improve performance under dynamic conditions, we incorporated an attention mask into the self-attention mechanism, restricting each frame’s receptive field to a local temporal window centered on that frame. We experimented with different window sizes and selected the optimal configuration for each speaker trajectory individually.

For instance, in the trajectory illustrated above, the first speaker achieved the best performance with a global (unmasked) attention window, whereas the second speaker reached the highest \ac{SISDR} with a 160-frame window.
On average, applying the attention mask led to approximately a 1~dB improvement in cases with large \ac{DOA} gaps, and even larger gains for scenarios involving smaller angular separations (see Table~\ref{tab:dynamic}). These results suggest that limiting temporal attention helps the model better handle frame-level variations in relative spatial positioning, thereby enhancing robustness in moving-speaker conditions.

\begin{table}[ht]
    \setlength{\abovecaptionskip}{0pt}
\setlength{\belowcaptionskip}{1pt}
\caption{SISDR\rev{ and \ac{STOI}} results for two moving speakers for 2 microphones\rev{. Each entry: \ac{SI-SDR} [dB] / \ac{STOI}}.}
\label{tab:dynamic}
\begin{center}
{\begin{tabular}{@{}lccc@{}}
\toprule
\textbf{model / test set} & (i) & (ii) & (iii) \\ \midrule
Unprocessed & -1.78/\rev{0.605} & -1.77/\rev{0.622} & -1.86/\rev{0.625} \\ \midrule
FasNet \cite{luo2019fasnet} & 4.30/\rev{0.727} & 2.71/\rev{0.707} & 5.70/\rev{0.765} \\
SpatialNet \cite{quan2024spatialnet} & 6.06/\rev{0.764} & 3.86/\rev{0.735} & 8.94/\rev{0.831} \\ \midrule
SpatialNet RPC & 6.83/\rev{0.771} & 3.60/\rev{0.712} & 8.91/\rev{0.825} \\
SpatialNet RPC oracle window & 7.90/\rev{0.776} & 6.10/\rev{0.686} & 9.98/\rev{0.816} \\
Unet RPC & 10.72/\rev{0.846} & 8.34/\rev{0.802} & 11.90/\rev{0.867} \\
Unet RPC oracle window & 11.63/\rev{0.859} & 10.01/\rev{0.822} & 12.96/\rev{0.885} \\
\bottomrule
\end{tabular}}
\label{table:results_moving}
\end{center}
\vspace{-20pt}
\end{table}



\begin{figure}[t]
    \centering
    \begin{subfigure}{0.48\columnwidth}
        \centering
        \includegraphics[width=\linewidth]{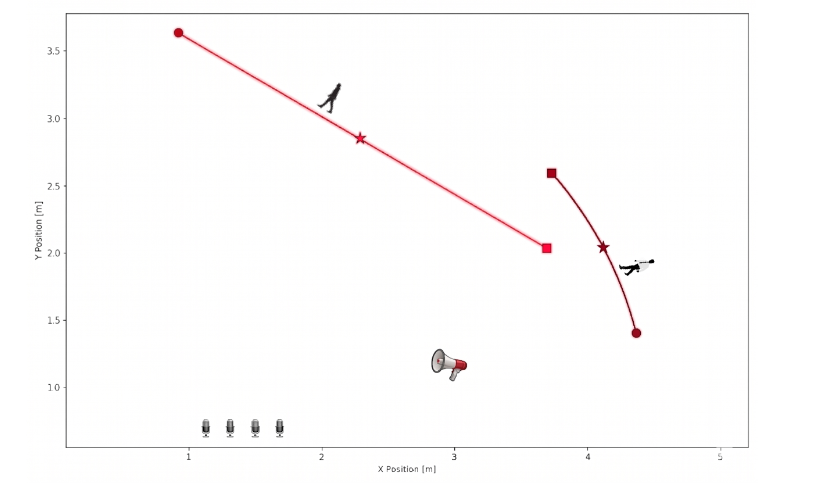}
        \caption{Speaker trajectory.}
        \label{fig:speaker-trajectory}
    \end{subfigure}
    \hfill
    \begin{subfigure}{0.48\columnwidth}
        \centering
        \includegraphics[width=\linewidth]{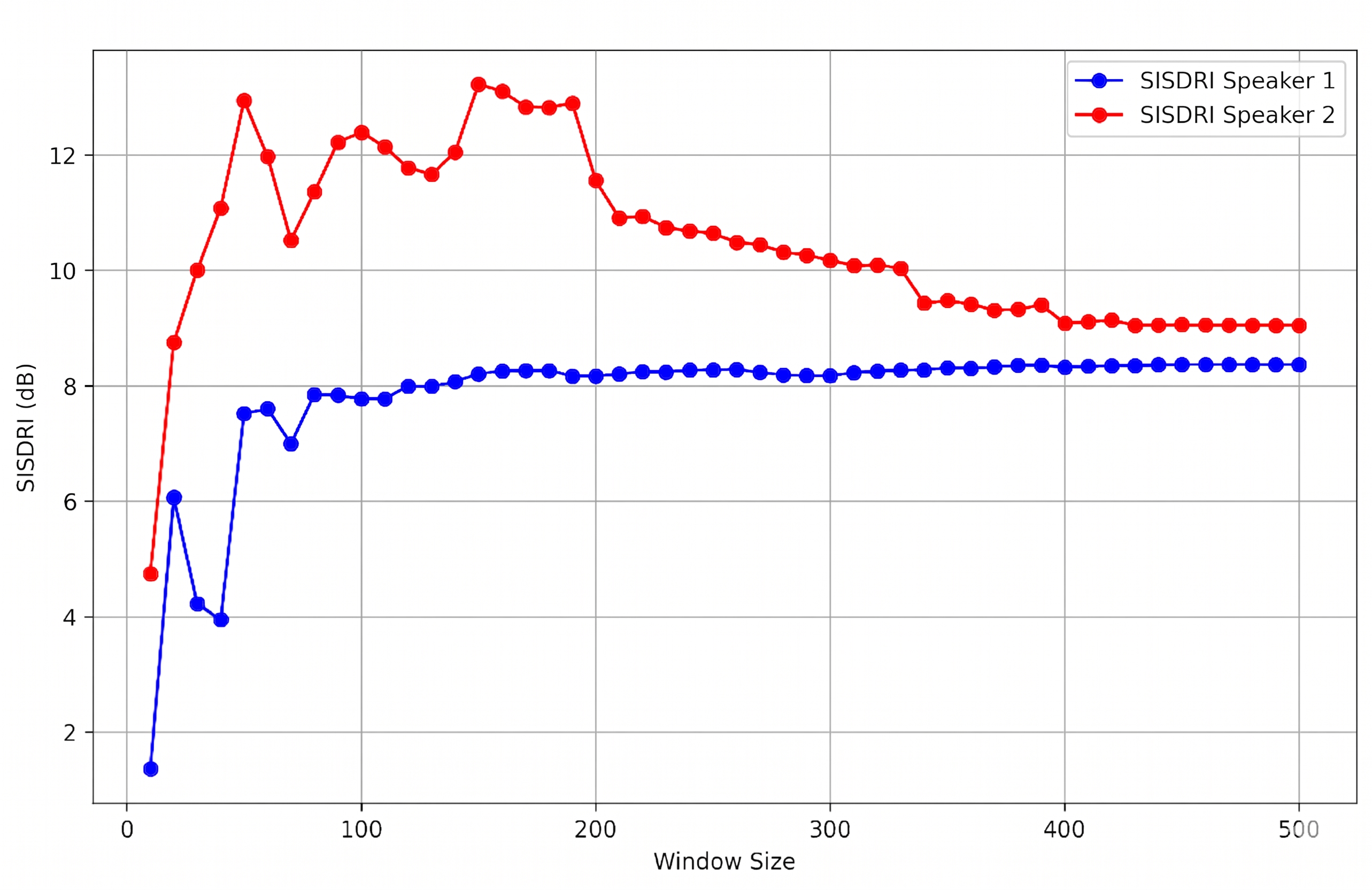}
        \caption{SI-SDRi vs. window size.}
        \label{fig:sisdri-vs-window-size}
    \end{subfigure}
        \setlength{\abovecaptionskip}{-3pt}
\setlength{\belowcaptionskip}{-2pt}
    \caption{(a) Speaker trajectory. (b) Example of SI-SDRi as a function of the window size.}
    \label{fig:trajectory-window}
\end{figure}

\section{\rev{Scope and Limitations}}
\rev{\label{sec:limitations}}
\rev{\textit{Overlapped speech and a known number of speakers.} The proposed method assumes that all $Q$ speakers are active throughout the processed segment and that $Q$ is known, since the command vocabulary depends on $Q$. In real conversations, however, overlap-free speech occupies most of the speaking time~\cite{ccetin2006analysis}, and the number of active speakers may change even within a single utterance. In single-speaker segments, the relative command is not defined, and in segments with a different number of speakers, the command must be re-specified. A practical system should therefore combine the \ac{RPC}-based extractor with a front-end that tracks the number of active speakers, e.g., a diarization or speaker-counting module, and apply the extractor only to overlapped segments, in the spirit of continuous speech separation~\cite{chen2020continuous}; overlap-free segments can be passed through a speech enhancement module.}

\rev{\textit{Changing relative positions.} The \ac{RPC} is meaningful as long as the relation between the speakers holds. If the speakers swap their relative positions within an utterance, the command becomes ambiguous; whether the model then follows the spatial relation or implicitly tracks the speaker identity is an interesting question for future research.}

\rev{\textit{Minimal spatial separation.} Configurations in which the speakers are neither angularly ($\phi\le10^\circ$) nor radially ($\psi\le0.5$~m) separated are not covered by the proposed cues.}

\section{Conclusions}
We presented a spatially guided \acf{TSE} framework that uses \ac{RPC} instead of explicit \ac{DOA} estimates. These cues represent discrete, interpretable spatial relationships (e.g., left/right and near/far) between concurrent speakers. This relative command is intuitive and semantically meaningful, and it is typically easier and more reliable to infer than a precise DOA, especially in reverberant or noisy environments. Across diverse datasets, microphone configurations, and mixture complexities, the approach consistently demonstrated robustness, including in scenarios with dynamic speakers. By aligning the conditioning signal with frame-level relational cues, the method maintains stable extraction under motion, offering a practical and interpretable alternative for multi-channel speaker extraction. \rev{We further showed that the improvements stem from selecting the commanded speaker rather than from noise suppression, and that the models exploit the relative configuration of the speakers rather than fixed regions of the room\rev{, and that models trained with directional noise only remain robust to diffuse babble noise}.}
\rev{Remaining challenges include handling overlap-free segments and a time-varying number of speakers, scaling the relative specification to more than three speakers, and robustness to nearby talkers.}

\bibliographystyle{IEEEtran}
\bibliography{main}

\vfill\pagebreak

\end{document}